\documentclass{article}

\usepackage{arXiv}

\usepackage[utf8]{inputenc} % allow utf-8 input
\usepackage[T1]{fontenc}    % use 8-bit T1 fonts
\usepackage{hyperref}       % hyperlinks
\usepackage{url}            % simple URL typesetting
\usepackage{booktabs}       % professional-quality tables
\usepackage{amsfonts}       % blackboard math symbols
\usepackage{nicefrac}       % compact symbols for 1/2, etc.
\usepackage{microtype}      % microtypography
\usepackage{mathrsfs}
\usepackage[table]{xcolor}         % colors

\hypersetup{
    colorlinks=true,
    linkcolor=blue
}
\usepackage{graphicx}       % figure
\usepackage{amsmath}        % equition
\usepackage{pifont}
\usepackage{subcaption}
\usepackage{tikz}
\usepackage{quantikz}
\usepackage{multirow}
\usepackage{array}
\usepackage{wrapfig}
\newcommand{\name}{SuperEncoder}

\newcommand{\squishlist}{
   \begin{list}{$\bullet$}
    { 
    \setlength{\itemsep}{0pt}      \setlength{\parsep}{0pt}
      \setlength{\topsep}{3pt}       \setlength{\partopsep}{0pt}
      \setlength{\listparindent}{-2pt}
      \setlength{\itemindent}{-5pt}
      \setlength{\leftmargin}{1em} \setlength{\labelwidth}{0em}
      \setlength{\labelsep}{0.5em} } }

\newcommand{\squishend}{
    \end{list}  }

\newcommand{\figref}[1]{Fig.~\ref{fig:#1}}
\title{\name{}: Towards Universal Neural Approximate Quantum State Preparation}

\author{
    Yilun Zhao\textsuperscript{\ref{equal_contrib}} \\
    Institute of Computing Technology \\
    Chinese Academy of Sciences\\
    \And
    Bingmeng Wang\textsuperscript{\ref{equal_contrib}} \\
    Capital Normal University \\
    \And
    Wenle Jiang\textsuperscript{\ref{equal_contrib}} \\
    Beijing University of Posts and Telecommunications \\
    \And
    Xiwei Pan\textsuperscript{\ref{equal_contrib}} \\
    University of Electronic Science and Technology of China \\
    \And
    Bing Li \\
    Capital Normal University \\
    \And
    Yinhe Han \\
    Institute of Computing Technology \\
    Chinese Academy of Sciences \\
    \And
    Ying Wang\thanks{Corresponding author. Email: wangying2009@ict.ac.cn} \\
    Institute of Computing Technology \\
    Chinese Academy of Sciences \\
}

\begin{document}

\begingroup\renewcommand\thefootnote{\textsection}
\footnotetext{\label{equal_contrib}Equal contributions.}
\endgroup

\maketitle

% \input{motivation}
% \input{design}
% \input{experiment}

%%%%%%%%%%%%%%%%%%%%%%%%%%% ABSTRACT %%%%%%%%%%%%%%%%%%%%%%%%%%%%%%
\begin{abstract}
  % The abstract paragraph should be indented \nicefrac{1}{2}~inch (3~picas) on
  % both the left- and right-hand margins. Use 10~point type, with a vertical
  % spacing (leading) of 11~points.  The word \textbf{Abstract} must be centered,
  % bold, and in point size 12. Two line spaces precede the abstract. The abstract
  % must be limited to one paragraph.
  %%%%%%___ A step towards classical ML-assisted general-purpose quantum state preparation.

  % Quantum State Preparation (QSP) stands for one of the major obstacles to achieving quantum advantage.
% Numerous quantum algorithms assume the classical data has been loaded into the quantum state, a process known as Quantum State Preparation (QSP).
Numerous quantum algorithms operate under the assumption that classical data has already been converted into quantum states,
a process termed Quantum State Preparation (QSP).
However, achieving precise QSP requires a circuit depth that scales exponentially with the number of qubits,
making it a substantial obstacle in harnessing quantum advantage.
Recent research suggests using a Parameterized Quantum Circuit (PQC) to approximate a target state, offering a more scalable solution with reduced circuit depth compared to precise QSP.
Despite this, the need for iterative updates of circuit parameters results in a lengthy runtime, limiting its practical application.
In this work, we demonstrate that it is possible to leverage a pre-trained neural network to directly generate the QSP circuit for arbitrary quantum state, thereby eliminating the significant overhead of online iterations.
Our study makes a steady step towards a universal neural designer for approximate QSP.
%To overcome this challenge, we introduce \name{}, a pre-trained classical neural network model designed to directly estimate the parameters of a PQC for any given quantum state.
%By eliminating the need for iterative parameter tuning,
%\name{} represents a pioneering step towards iteration-free approximate QSP.

\end{abstract}
%%%%%%%%%%%%%%%%%%%%%%%%%%% ABSTRACT %%%%%%%%%%%%%%%%%%%%%%%%%%%%%%

\section{Introduction}

% \todo{QC and its progress}
Quantum Computing (QC) leverages quantum mechanics principles to address classically intractable problems~\cite{shor1999polynomial,nielsen2010qc-and-qi}.
% With the rapid development of hardware~\cite{monz201114,IBM1121,arute2019google-supremacy},
% there have been multiple demonstrations of quantum computing's ability to address practical challenges across different fields, including finance~\cite{herman2023quantum_computing_for_finance}, physics~\cite{kim2023evidence_before_advantage}, and artificial intelligence (AI)~\cite{jiang2021quantumflow_ncomm}\todo{more citations}.
% QC has shown its practical utility across different fields~\cite{herman2023quantum_computing_for_finance,kim2023evidence_before_advantage,biamonte2017qml_nature_review,jiang2021quantumflow_ncomm}.
% \todo{why QSP?}
Various quantum algorithms have been developed, encompassing
% Quantum State Preparation (QSP) is a crucial subroutine in various quantum algorithms, including
quantum-enhanced linear algebra~\cite{Harrow2009-ai_hhl,srinivasan2018learning_infer_hilbert_nips,schuld2016prediction_linear_regress_algebra},
Quantum Machine Learning (QML)~\cite{li2022concentration_AG,jiang2021quantumflow_ncomm,abbas2021power_of_qnn,mitarai2018quantum_circuit_learning,tian2023recent_qnn_survey,bausch2020recurrent_qnn},
quantum-enhanced partial differential equation solvers~\cite{lubasch2020vqa_for_nonlinear_problems,gonzalez2021simulate_option_price_PDE}, etc.
A notable caveat is that those algorithms assume that classical data has been efficiently loaded into a specific quantum state,
% A notable caveat of these algorithms is that they assume classical data has been loaded into a specific quantum state,
% giving rise to the problem of Quantum State Preparation (QSP).
a process known as Quantum State Preparation (QSP).

% \todo{We don't need to talk about angle encoding if our story is QSP?}

% An essential subroutine in quantum algorithms is Quantum State Preparation (QSP), \todo{definition}.
% Quantum State Preparation (QSP), an essential subroutine in quantum computing (QC), enables setting up the system's initial state.
% This step is vital for various applications, including generating codewords for quantum error correction,
% This step is vital for various applications, including loading classical data into quantum states in Quantum Machine Learning (QML)~\cite{li2022concentration_AG}\todo{more citations},
% loading initial conditions to solve Partial Differential Equations (PDE),
% and \todo{linear algorithm, grover's algorithm, etc}.

% \todo{Applications of QSP~\cite{Harrow2009-ai_hhl,srinivasan2018learning_infer_hilbert_nips,biamonte2017qml_nature_review}}

% \todo{QSP (AE)~\cite{mottonen2004pennylane_AME,plesch2011qsp_PRA}}

% \todo{AE}
% The problem of QSP has long been investigated~\cite{mottonen2004pennylane_AME,plesch2011qsp_PRA}.
However, the realization of QSP presents significant challenges.
Ideally, we expect each element of the classical data to be precisely transformed into an amplitude of the corresponding quantum state.
This precise QSP is also known as Amplitude Encoding (AE).
% Precise QSP involves encoding each element of the classical data into an amplitude of corresponding quantum state, a.k.a. Amplitude Encoding (AE),
% which has long been investigated~\cite{mottonen2004pennylane_AME,plesch2011qsp_PRA,long2001efficient_qsp,shende2005synthesis_qsp}.
However,
a critical yet unresolved problem of AE is that the required circuit depth grows exponentially with respect to the number of qubits~\cite{mottonen2004pennylane_AME,plesch2011qsp_PRA,long2001efficient_qsp,shende2005synthesis_qsp,sun2023asymptotically_optimal_qsp}.
% It is well-known that AE requires a quantum circuit with exponential depth with the number of qubits.
% It is well-known that AE requires a quantum circuit with exponential depth with the number of qubits.
Extensive efforts have been made to alleviate this issue, but they fail to address it fundamentally.
For example, while some methods introduce ancillary qubits for shallower circuit~\cite{zhao2019state_ancilla_qpe,zhang2021low_depth_QSP_ancilla,araujo2023configurable_sub_qip_qsp},
they may encounter an exponential number of ancillary qubits.
Other methods aim at preparing \emph{special} quantum states with lower circuit depth, being only effective for either sparse states~\cite{gleinig2021efficient_special_sparse_qsp,mao2024towards_sparse_special_qsp}
or states with some special distributions~\cite{gonzalez2024efficient_special_qsp,iaconis2024qsp_normal_distribution}.
% Other methods aims at preparing \emph{special} states with low depth~\cite{gonzalez2024efficient_special_qsp,gleinig2021efficient_special_sparse_qsp,mao2024towards_sparse_special_qsp,iaconis2024qsp_normal_distribution}.
To summarize,
realizing AE for \emph{arbitrary} quantum states still remains \emph{non-scalable} due to its exponential resource requirement with respect to the number of qubits.
Moreover, in the Noisy Intermediate-Scale Quantum (NISQ) era~\cite{preskill2018quantum},
hardware has limited qubit lifetimes and confronts a high risk of decoherence errors when executing deep circuits,
further exacerbating the problem of AE.

% the most straightforward way is Amplitude Encoding (AE), \todo{definition, more related work}.
% However, the number quantum gates\todo{pointer to section}\todo{gate num of depth?} required to implement AE for arbitrary quantum states grows exponentially with increasing number of qubits.

% \todo{AAE}
In fact, precise QSP is unrealistic in the present NISQ era due to the inherent errors of quantum devices.
Hence, iteration-based Approximate Amplitude Encoding (AAE) emerges as a promising technique~\cite{zoufal2019quantum_gan_npjq,nakaji2022aae_pr_research,wang2023robuststate}.
% Fortunately, there are many application scenarios where precise QSP is not mandatory.
% Therefore, Approximate Amplitude Encoding (AAE) is proposed and  explored~\cite{zoufal2019quantum_gan_npjq,nakaji2022aae_pr_research,wang2023robuststate},
Specifically,
AAE constructs a quantum circuit with tunable parameters,
then it iteratively updates the parameters to approximate a target quantum state.
Since the updating of parameters can be guided by states obtained from noisy devices,
AAE is robust to noises, becoming especially suitable for NISQ applications.
% which approximates a target state via iteratively tuning a Parameterized Quantum Circuit (PQC) (\ssecref{qsp}).
More importantly,
% AAE has been proven to have polynomial circuit depth~\cite{nakaji2022aae_pr_research}, rendering it a more scalable approach than AE.
AAE has been shown to have shallow circuit depth~\cite{nakaji2022aae_pr_research,wang2023robuststate}, making it more scalable than AE.

% In some application scenarios, precise preparation of target state is not required.
% For example, \todo{remove?}Angle Encoding (AGE) (pointer to section) is often adopted for QML tasks \todo{implicit transformation}.
% AGE possesses simple circuit structure and requires only linear circuit depth, thus it has been widely implemented as the default encoding module in QML frameworks.
% However, AGE can be much less effective than AE and dampens system performance, as we shown in \todo{seciton xx}.
% To reduce the circuit depth of AE, Approximate Amplitude Encoding has been proposed and explored~\cite{zoufal2019quantum_gan_npjq,nakaji2022aae_pr_research},
% which essentially \emph{trains} a Variational Quantum Circuit\todo{citations and point to section} to approximate target quantum state.
%\begin{wrapfigure}{r}{0.4\textwidth}
\begin{figure}[h]
    \centering
    \includegraphics[width=0.4\textwidth]{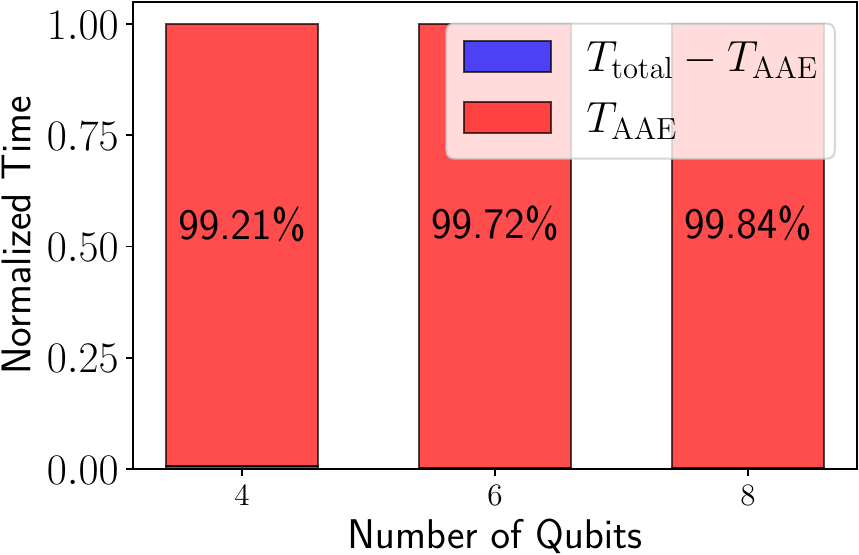}
    \caption{Breakdown of normalized runtime for QNN inference. Original data are listed in Table~\ref{tab:aae_breakdown}.}
    \label{fig:breakdown_qnn_infer}
\end{figure}
%\end{wrapfigure}

% \todo{limitations}
Unfortunately, AAE possesses a drawback that significantly undermines its potential advantages --- the lengthy runtime stemming from iterative optimizations of parameters.
% Although AAE has been shown to be both accurate and resource efficient (shallow circuit depth), its long runtime prohibits the usage in practical scenarios.
% That is, for any given target state, we must train a quantum circuit
% For example, when a QML model is trained and deployed, performing online iterations for each input data during the inference time is impractical.
For example, when a Quantum Neural Network (QNN)~\cite{bausch2020recurrent_qnn} is trained and deployed, the runtime of AAE dominates the inference time as we demonstrated in Fig.~\ref{fig:breakdown_qnn_infer}.
Since loading classical data into quantum states becomes the bottleneck,
the potential advantage of QNN diminishes no matter how efficient the computations are done on quantum devices.
% We ask, if most of the time is spent on loading classical data into quantum states,
% where lies the advantage of QML?
% which would otherwise significantly dampen the expected quantum advantage (Section~\ref{ssec:motivation}).
% \todo{envision a situation in QML}.

Compared to AAE, AE employs a pre-defined arithmetic decomposition procedure to construct a circuit, thereby becoming much \emph{faster} than AAE at runtime.
Therefore, it is natural to ask: can we realize both \emph{fast} and \emph{scalable} methods for \emph{arbitrary} QSP?
This is precisely the question we tackle in this paper.
Overall, we present three major contributions.
% In this work, we address a fundamental and challenging question:
% how to realize \emph{fast}, \todo{\emph{accurate?}} and \emph{scalable} \emph{arbitrary} QSP?
% how to develop a \emph{fast} and \emph{scalable} methodology for arbitrary QSP?

\squishlist{}
    % To our best knowledge, this is the first time
    \item
    % Given an arbitrary quantum state $|\boldsymbol{d}\rangle$,
    Given a Parameterized Quantum Circuit  (PQC) $U(\boldsymbol{\theta})$ that approximates a target quantum state, with $\boldsymbol{\theta}$ the parameter vector.
    We show that there exists a \emph{deterministic} transformation $f$ that could map an arbitrary state $|\boldsymbol{d}\rangle$ to its corresponding parameters $\boldsymbol{\theta}$.
    Consequently, the parameters can be designated by $f$ without time-intensive iterations.
    \item We show that the mapping $f$ is \emph{learnable} by utilizing a classical neural network model, which we term as \name{}.
    With \name{}, you can have your cake and eat it too, i.e., simultaneously realizing \emph{fast} and \emph{scalable} QSP.
    We develop a prototype model and shed light on insights into its training methodology.
    \item We verify the effectiveness of \name{} on both synthetic dataset and representative downstream tasks, paving the way toward iteration-free approximate quantum state preparation.

\squishend{}

\vspace{-5pt}
\section{Preliminaries}

In this section, we commence with some basic concepts about quantum computing~\cite{nielsen2010qc-and-qi},
and then proceed to a brief retrospect of existing QSP methods. 

\subsection{Quantum Computation}

%-------- what is qubit, what is state, what is gate. Describe how $\theta$ impacts qubit state in bloch sphere.

% Quantum mechanics operates in the Hilbert space $\mathcal{H}$

We use Dirac notation throughout this paper. A \emph{pure quantum state} is defined by a vector $|\cdot\rangle$ named `ket', with the unit length.
A state can be written as $|\psi\rangle = \sum_{j=1}^{N} \alpha_j |j\rangle$ with $\sum_{j} |\alpha_j|^2 = 1$, 
where $|j\rangle$ denotes a computational basis state
and $N$ represents the dimension of the complex vector space.
\emph{Density operators} describe more general quantum states.
Given a mixture of $m$ pure states $\{|\psi_i\rangle\}_{i=1}^{m}$ with probabilities $p_i$ and $\sum_{i}^{m} p_i = 1$,
the density operator $\rho$ denotes the \emph{mixed state} as
$\rho = \sum_{i=1}^{m} p_i | \psi_i\rangle \langle \psi_i |$ with $\mathrm{Tr}(\rho) = 1$,
where $\langle \cdot |$ refers to the conjugate transpose of $|\cdot\rangle$.
Generally, we use the term \emph{fidelity} to describe the similarity between an erroneous quantum state and its corresponding correct state.

The fundamental unit of quantum computation is the quantum bit, or \emph{qubit}.
% Quantum information is typically described using the quantum states of these qubits.
A qubit's state can be expressed as $\psi = \alpha |0\rangle + \beta |1\rangle$.  
Given $n$ qubits, the state is generalized to 
$|\psi\rangle = \sum_{j}^{2^n}|j\rangle$,
where $|j\rangle = |j_1 j_2 \cdots j_n\rangle$ 
with $j_k$ the state of $k$th qubit in computational basis, and $j = \sum_{k=1}^{n}2^{n-k} j_k$.
% In quantum computing, these states undergo manipulation through various quantum operations.
% A quantum state evolves from one to another state through \emph{quantum operations}, including quantum gates and measurements.
Applying \emph{quantum operations} evolves a system from one state to another. 
Generally, these operations can be categorized into quantum gates and measurements.
Typical single-qubit gates include the Pauli gates
% \todo{Dirac notation}
% The quantum state is manipulated by quantum operations.
% Typical single-qubit operations include Pauli gates, 
$X\equiv \left[\begin{smallmatrix} 0 & 1 \\ 1 & 0 \end{smallmatrix} \right]$, 
$Y\equiv \left[\begin{smallmatrix} 0 & -i \\ i & 0 \end{smallmatrix} \right]$,
$Z\equiv \left[\begin{smallmatrix} 1 & 0 \\ 0 & -1 \end{smallmatrix} \right]$.
These gates have associated rotation operations 
$R_P(\theta)\equiv \mathrm{e}^{-i\theta P/2}$,
where $\theta$ is the rotation angle and $P\in\{X,Y,Z\}$\footnote{In this paper, $R_z,R_y$ are equivalent to $R_Z,R_Y$.}.
Muti-qubit operations create \emph{entanglement} between qubits, allowing one qubit to interfere with others.
In this work, we focus on the controlled-NOT (CNOT) gate, with the mathematical form of 
$\mathrm{CNOT}\equiv |0\rangle\langle0|\otimes\mathbf{I}_2 + |1\rangle\langle1|\otimes X$.
% To get classical information from quantum state, one needs to perform quantum measurements.
Quantum measurements extract classical information from quantum states, which is described by a collection $\{M_m\}$ with $\sum_m M_m^\dagger M_m = \mathbf{I}$.
Here, $m$ refers to the measurement outcomes that may occur in the experiment, with a probability of 
$p(m) = \langle\psi|M_m^\dagger M_m|\psi\rangle$.
The post-measurement state of the system becomes
$M_m|\psi\rangle/p(m)$.

\begin{figure}[ht!]
    \centering
    \scalebox{0.8}{
    \begin{quantikz}
        \lstick{$|0\rangle$} 
        & \gate{R_{y}(\theta_0)}
        \gategroup[wires=4,steps=3,style={inner sep=3pt}, label style={label position=below, anchor=north, yshift=-0.2cm}]{Block \# 0}  
        \gategroup[wires=4,steps=1,style={dashed, inner sep=0pt, fill=blue!20},background]{}
        % \lstick{$|0\rangle$} & \gate{R_{y}(\theta_0)}\gategroup[wires=4,steps=4]{}  
        & \ctrl{1}
        \gategroup[wires=4,steps=2,style={dashed, inner sep=0pt, fill=red!20},background]{}
        & \qw  & \qw
        & \gate{R_{y}(\theta_4)} 
        \gategroup[wires=4,steps=3,style={inner sep=3pt}, label style={label position=below, anchor=north, yshift=-0.2cm}]{Block \# 1}  
        & \ctrl{1} & \qw       & \qw      & \qw \\
        \lstick{$|0\rangle$} & \gate{R_{y}(\theta_1)}                               
        & \targ{} & \ctrl{1}     & \qw
        & \gate{R_{y}(\theta_5)} 
        & \targ{}  & \ctrl{1}  & \qw & \qw \\
        \lstick{$|0\rangle$} & \gate{R_{y}(\theta_2)}                               
        & \ctrl{1} & \targ{}    & \qw 
        & \gate{R_{y}(\theta_6)} 
        & \ctrl{1}      & \targ{}   & \qw & \qw \\
        \lstick{$|0\rangle$} & \gate{R_{y}(\theta_3)}                               
        & \targ{} & \qw        & \qw 
        & \gate{R_{y}(\theta_7)} 
        & \targ{}      & \qw       & \qw
        \slice{Approximated state of $|\boldsymbol{d}\rangle$}  & \qw 
    \end{quantikz} 
    }
    \caption{An example PQC with two blocks, with each block consisting of a rotation layer (filled blue) plus an entangler layer (filled red).}
    \label{fig:example_pqc}
\end{figure}
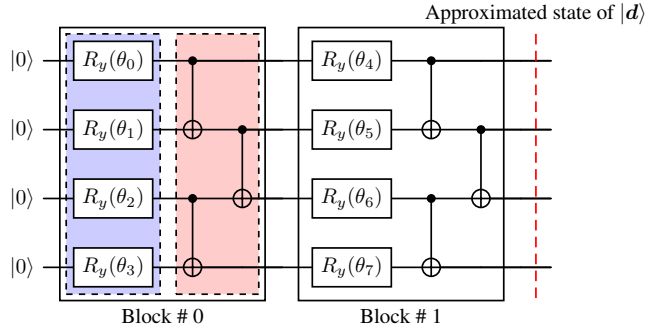

A \emph{quantum circuit} is the graphical representation of a series of quantum operations, which can be mathematically represented by a unitary matrix $U$.
In the NISQ era, PQC plays an important role as it underpins variational quantum algorithms~\cite{farhi2014quantum_qaoa,vqe_nature_comm}.
Typical PQC has the form of $U(\boldsymbol{\theta}) = \prod_i U_i(\theta_i) V_i$, 
where $\boldsymbol{\theta}$ is its parameter vector, $U_i(\theta_i) = \mathrm{e}^{-i\theta_i P_i / 2}$ with $P_i$ denoting a Pauli gate, 
and $V_i$ denotes a fixed gate such as CNOT.
For example, a PQC composed of $R_y$ gates and CNOT gates is depicted in \figref{example_pqc}.
% depicts a sample PQC composed of .
% An example PQC is shown in \figref{example_pqc}.

% \todo{RX RY...}
% \todo{bloch sphere}

\subsection{Quantum State Preparation}
\label{ssec:qsp}

% - mottonen algorithm

% \todo{an example circuit for mottonen}

% - variational QSP.

% - Formalize QSP.
Successful execution of many quantum algorithms requires an initial step of loading classical data into a quantum state~\cite{biamonte2017qml_nature_review,Harrow2009-ai_hhl}, 
a process known as \emph{quantum state preparation}.
This procedure involves implementing a quantum circuit to evolve a system to a designated state. 
Here, we focus on \emph{amplitude encoding} and formalize its procedure as follows.
% a $n$-qubit state from $|0\rangle^{\otimes n}$ to an arbitrary state $|\psi\rangle$.
% In this paper, we focus on the application scenarios of QSP in quantum-enhanced linear algebra~\cite{Harrow2009-ai_hhl} and QML~\cite{biamonte2017qml_nature_review}\todo{citations refresh}.
% These applications require to load classical data into quantum state.
% Then the problem of QSP in this context can be described as: 
% The most straightforward way of precise QSP is \emph{amplitude encoding} (AE).
% Let $\boldsymbol{d}$ be an $N$-dimensional classical vector,
Let $\boldsymbol{d}$ be a real-valued $N$-dimensional classical vector, 
AE encodes $\boldsymbol{d}$ into the amplitudes of an $n$-qubit quantum state $|\boldsymbol{d}\rangle$,
where $N=2^n$.
More specifically, the data quantum state is represented by $|\boldsymbol{d}\rangle = \sum_{j=0}^{N-1} d_j |j\rangle$,
% \begin{equation}
%     |\boldsymbol{d}\rangle = \sum_{j=0}^{N-1} d_j |j\rangle,
% \end{equation}
where $d_j$ denotes the $j$th element of the vector $\boldsymbol{d}$, and $|j\rangle$ refers to a computational basis state.
% Given an $n$-qubit ($N=2^n$) system with initial state $|0\rangle^{\otimes n}$, 
The main objective is to generate a quantum circuit $U$
that initializes an $n$-qubit system by $U |0\rangle^{\otimes n} = \sum_{j=0}^{N-1} \alpha_j|j\rangle$,
% \begin{equation}
%     U |0\rangle^{\otimes n} = \sum_{j=0}^{N-1} \alpha_j|j\rangle,
% \end{equation}
whose amplitudes $\{\alpha_j\}$ are equal to $\{d_j\}$.
It is widely recognized that constructing such a circuit generally necessitates a circuit depth that scales exponentially with $n$~\cite{mottonen2004pennylane_AME,plesch2011qsp_PRA}.
% making AE impractical, particularly in current NISQ era.
This property makes AE impractical in current NISQ era, 
as decoherence errors~\cite{krantz2019quantum_eng_guide_sc} can severely dampen the effectiveness of AE as the number of qubits increases~\cite{wang2023robuststate}.
% The presence of noises brings up the opportunity

% Although precise QSP is hard to realize under the presence of noises, 
% Conversely, approximate QSP 
% Fortunately, many quantum algorithms are compatible with \emph{approximate} QSP~\cite{zoufal2019quantum_gan_npjq}.

In response to the inherent noisy nature of current devices,
\emph{approximate amplitude encoding} has emerged as a promising technique~\cite{zoufal2019quantum_gan_npjq,nakaji2022aae_pr_research,wang2023robuststate}.
% Taking this opportunity, \emph{approximate amplitude encoding} has emerged as a promising technique~\cite{zoufal2019quantum_gan_npjq,nakaji2022aae_pr_research,wang2023robuststate}.
% Specifically, AAE utilizes a PQC $U(\boldsymbol{\theta})$ (a.k.a. ansatz) to approximate the target quantum state,
Specifically, AAE utilizes a PQC (a.k.a. ansatz) to approximate the target quantum state by iteratively updating the parameters of circuit, 
following a similar procedure of other variational quantum algorithms~\cite{vqe_nature_comm,farhi2014quantum_qaoa}.
AAE has been shown to be more advantageous for NISQ devices due to its ability to mitigate coherent errors through flexible adjustment of circuit parameters, coupled with its lower circuit depth~\cite{wang2023robuststate}.
% The PQC is also known as \emph{ansatz}, which we denote by $U(\boldsymbol{\theta})$
We denote an ansatz as $U(\boldsymbol{\theta})$, where
$\boldsymbol{\theta}$ refers to a vector of tunable parameters for optimizations.
A typical ansatz consists of several blocks of operations with the same structure.
For example, a two-block ansatz with 4 qubits is shown in \figref{example_pqc},
where the rotation layer is composed of single-qubit rotational gates $R_y(\theta_r) = \mathrm{e}^{-i\theta_r Y/2}$,
and the entangler layer comprises CNOT gates.
Note that the entangler layer is configurable and hardware-native, 
which means that we can apply CNOT gates to physically adjacent qubits,
% Note that the connections of CNOT gates can be configured to be \emph{native} to target hardware, 
thereby eliminating the necessity of additional SWAP gates to overcome the topological constraints~\cite{li2019tackling}.
This type of PQC is also known as \emph{hardware-efficient ansatz}~\cite{nature_hw_efficient_ansatz}, 
being widely adopted in previous studies of AAE~\cite{zoufal2019quantum_gan_npjq,nakaji2022aae_pr_research,wang2023robuststate}.

\section{\name{}}

\subsection{Motivation}

\label{ssec:motivation}

% - Algorithmic analysis is more than enough, i.e., the complexity of different QSP methods.

\begin{figure}[h!]
    \centering
    \begin{subfigure}{0.75\textwidth}
        \centering
        \includegraphics[width=\linewidth]{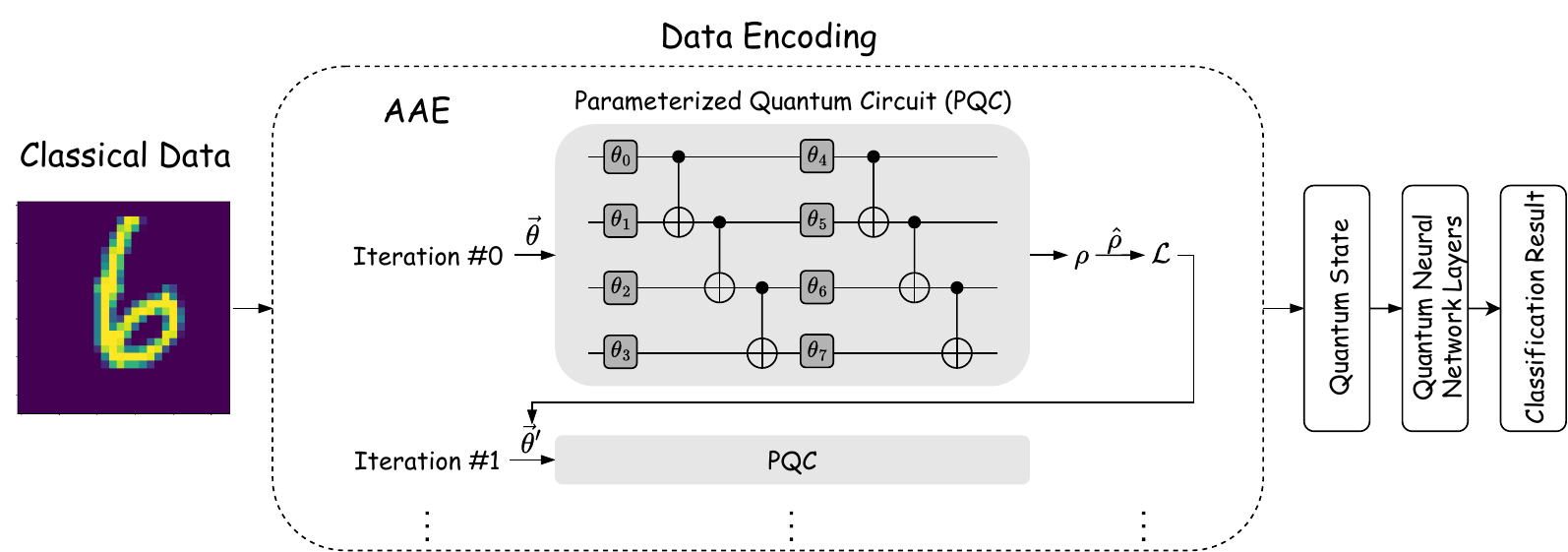}
        \caption{Inference process of AAE.}
        \label{fig:aae_workflow}
    \end{subfigure}
    \begin{subfigure}{0.75\textwidth}
        \centering
        \includegraphics[width=\linewidth]{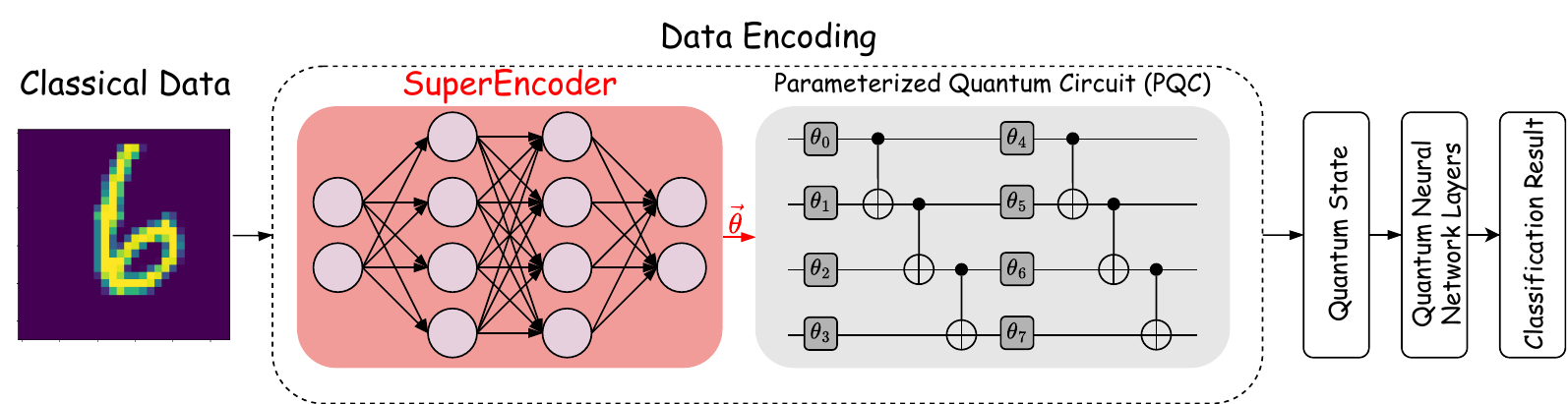}
        \caption{Inference process of \name{}. }
        \label{fig:superencoder_workflow}
    \end{subfigure}
    \caption{Comparison between AAE and \name{}.}
    \label{fig:overview}
\end{figure}

% In contrast to precise AE, AAE could realize high fidelity state preparation with $O(\text{poly}(n))$ circuit depth as proved by previous work~\cite{nakaji2022aae_pr_research}.
Although AAE can potentially realize high fidelity QSP with $O(\text{poly}(n))$ circuit depth~\cite{nakaji2022aae_pr_research}
with $n$ the number of qubits,
it requires repetitive \emph{online} tuning of parameters to approximate the target state, which may result in an excessively long runtime that undermines its feasibility.
% However, AAE requires repetitive \emph{online} tuning of parameters to approximate the target state, which may result in excessively long runtime that hinders its practical usage.
Specifically, we could consider a simple application scenario in QML.
The workflow with AAE is depicted in Fig.~\ref{fig:aae_workflow}.
During the inference stage, we must iteratively update the parameters of the AAE ansatz for each input classical data vector, 
which may greatly dampen the performance.
To quantify this impact, we measure the runtime of AAE-based data loading and the total runtime of model inference.
As one can observe from Table~\ref{tab:aae_breakdown},
AAE dominates the runtime, thereby becoming the performance bottleneck.

\begin{table}[h]
    \centering
    \begin{tabular}{|c|c|c|}
        \hline
        \rowcolor{lightgray}
        $n$ & $T_{\text{AAE}}$ (s) & $T_{\text{total}} - T_{\text{AAE}}$ (s) \\
        \hline
        4 & \textbf{5.0086} & 0.0397 \\
        6 & \textbf{20.1810} & 0.0573 \\
        8 & \textbf{59.4193} & 0.0978 \\
        \hline
    \end{tabular}
    \caption{\textbf{Performance overhead of AAE}. We break down the averaged inference runtime per sample from the MNIST dataset. 
    $T_{\text{AAE}}$ denotes time spent on loading classical data into quantum state using AAE, and $T_{\text{total}}$ refers to total runtime.}
    \label{tab:aae_breakdown}
\end{table}

The necessity of time-intensive iterations is grounded in the following assumption --- 
Given an arbitrary quantum state $|\psi\rangle$, 
there \emph{does not} exist a deterministic transformation 
$f: |\psi\rangle \rightarrow \boldsymbol{\theta},$ 
where $\boldsymbol{\theta}$ refers to the vector of parameters enabling a PQC to prepare an approximated state of $|\psi\rangle$.
% there is \emph{no deterministic transformation} $f(|\psi\rangle)$, \red{or we can say there is a data-dependent transformation $f_{\psi}:|\psi\rangle \rightarrow \boldsymbol{\theta}_{\psi},$ where $\boldsymbol{\theta}_{\psi}$ is parameters of PQC for data $|\psi\rangle$}
% whose output is exactly the required parameter vector $|\boldsymbol{\theta}\rangle$ that approximates $|\psi\rangle$.
This assumption seems intuitively correct given the randomness of target states.
However, we argue that a universal mapping $f$ exists for any arbitrary data state $|\psi\rangle$.
% However, we argue that a deterministic transformation may exist.\red{However, we argue that the transformation is data-free and there is a universal mapping $f:|\psi\rangle \rightarrow \boldsymbol{\theta}_{\psi},$ for arbitrary data states $|\psi\rangle$}
% The insight is that, there exists a deterministic procedure to generate a quantum circuit
Taking a little thought of AE, we see that it implies the following conclusion: 
given an arbitrary state $|\psi\rangle$, 
there exists an universal arithmetic decomposition procedure $g: |\psi\rangle \rightarrow U$
satisfying $U|0\rangle = |\psi\rangle$.
% Looking back at AE, 
% the corresponding circuit for an arbitrary state is generated using the same arithmetic decomposition procedure designed based on human expertise.
% a deterministic transformation exists between arbitrary states and their corresponding QSP circuit, albeit with non-scalable depth.
% This transformation is done by arithmetic decomposition based on human expertise.
Inspired by this deterministic transformation,
it is natural to ask: 
is there an universal transformation 
$g^\prime: |\psi\rangle \rightarrow U^\prime$ satisfying $E(U^\prime|0\rangle,|\psi\rangle) \leq \epsilon$?
Here $E$ denotes the deviation between the prepared state by a circuit $U^\prime$ and the target state, 
and $\epsilon$ refers to certain acceptable error threshold.
% Inspired by this deterministic transformation in AE, we advocate for exploring the relationship between an arbitrary state $|\psi\rangle$ and its associated $\boldsymbol{\theta}$.
% If a fixed transformation exists, there is no need to train a PQC for appro
Since the structure of PQC in AAE is the same for any target state,
$U^\prime$ is determined by $\boldsymbol{\theta}$.
Then, the problem is reduced to exploring the existence of 
$f: |\psi\rangle \rightarrow \boldsymbol{\theta}$.
Should $f$ exist, the overhead of online iterations could be eliminated, resulting in a novel QSP method being both fast and scalable.

% \todo{Table}
% Table~{\todo{xx}} shows the runtime comparison between AAE and AE, as well as the occupation of the time for training AE within the end-to-end inference runtime.

% \todo{Figure}

% \subsection{Legacy\todo{refine}}

% \todo{refresh}
% \subsection{}

% - Pre-training

% In light of the necessity of training $U(\boldsymbol{\theta})$ for each 
% To eradicate the overhead of AAE, we take  
% In this work, we take a fresh look and PQC-based approximate QSP.

% To eradicate the overhead of AAE, we design \name{}.

% ----- The primary goal of this paper is to 
% ----- Particularly, we ask the following question: 
% ----- Is the relationship between $|\mathcal{D}\rangle$ and  $\boldsymbol{\theta}$ \emph{deterministic}?
% ----- If so, there exists a fixed transformation $f$ such that $\boldsymbol{\theta}$ could be given by $\boldsymbol{\theta} = f(|\mathcal{D}\rangle)$.
% ----- Thus for an arbitrary data state $|\mathcal{D}\rangle$, we could approximate it by $U(f(|\mathcal{D}\rangle))$.

% \subsection{Learning Objective}

\subsection{Design Methodology}

\label{ssec:design}

Let $|\psi\rangle$ be the target state, and $U(\boldsymbol{\theta})$ be the PQC used in AAE with $\boldsymbol{\theta}$ the optimized parameters.
Our goal is to develop a model, termed \name{},
to approximate the mapping $f: |\psi\rangle \rightarrow \boldsymbol{\theta}$.
Referring back to the scenario in QML, the workflow with \name{} becomes iteration-free, as depicted in Fig.~\ref{fig:superencoder_workflow}.

Since neural networks could be used to approximate any continuous function~\cite{operatorapproximation_op_learn},
a natural solution is to use a neural network to approximate $f$.
Specifically, we adopt a Multi-Layer Perceptron (MLP) as the backbone model for approximating $f$.
However, training this model is nontrivial.
Particularly, we find it challenging to design a proper loss function.
In the remainder of this section, we explore three different designs and analyze their performance.

\begin{figure}[h!]
    \centering

    \begin{subfigure}{0.2\textwidth}
       \includegraphics[width=0.99\linewidth]{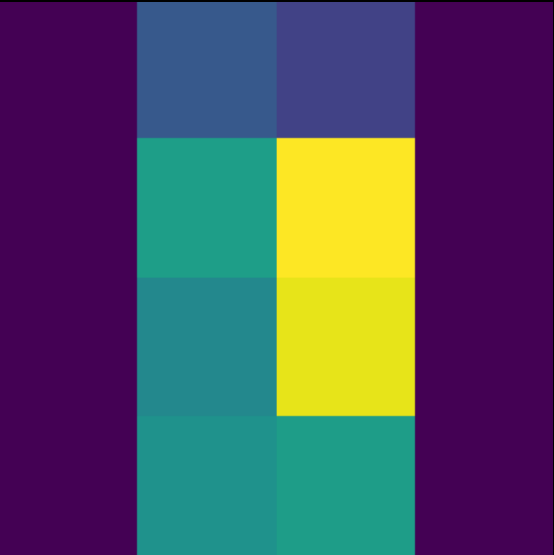}
       \caption{Target state.}
    \end{subfigure}
    \begin{subfigure}{0.2\textwidth}
       \includegraphics[width=0.99\linewidth]{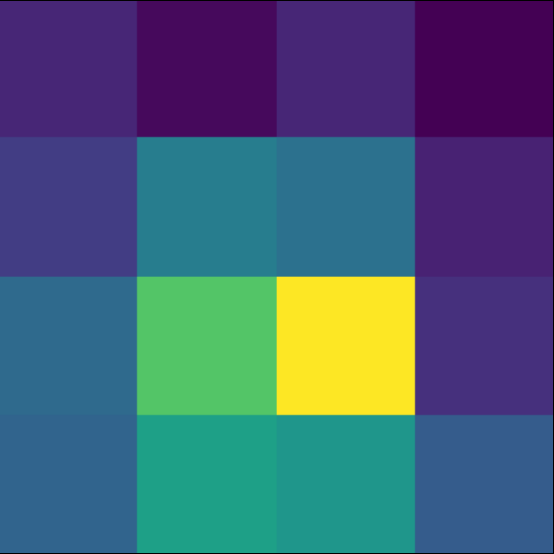}
       \caption{\name{}-$\mathcal{L}_1$}
    \end{subfigure}
    \begin{subfigure}{0.2\textwidth}
       \includegraphics[width=0.99\linewidth]{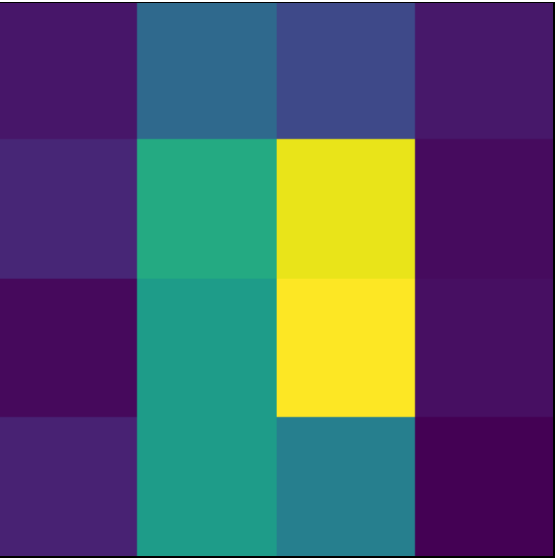}
       \caption{\name{}-$\mathcal{L}_3$}
    \end{subfigure}
    \caption{Virtualization of states generated by \name{} trained with different loss functions. $\mathcal{L}_2$ is omitted as it produces very similar results to $\mathcal{L}_3$.}
    \label{fig:virtual_prep_states_losses}
\end{figure}

The first and most straightforward method is \emph{parameter-oriented} training ---
setting the loss function $\mathcal{L}_1$ as the MSE between the target parameters
$\boldsymbol{\theta}$ from AAE and the output parameters $\boldsymbol{\hat{\theta}}$ from \name{}.
To evaluate the performance of $\mathcal{L}_1$, we train a \name{} using MNIST dataset,
and test if it could load a test digit image into a quantum state with high fidelity.
All images are downsampled and normalized into 4-qubit states for quick evaluation.
% As an initial exploration, we evaluate the performance of \name{} by applying it to a binary classification task on MNIST dataset and compare the classification accuracy with AAE.
% We adopt a simple MLP as the underlying network architecture, the detailed configuration is illustrated in Section~\ref{ssec:setup}.
% Let $f(\mathbf{W},|\boldsymbol{d}\rangle)$ be the MLP with $\mathbf{W}$ the model weights.
% The output is given by $\boldsymbol{\hat{\theta}} = f(\mathbf{W},|\boldsymbol{d}\rangle)$.

% \begin{table}[h]
%\begin{wraptable}{r}{0.35\textwidth}
\begin{table}[h]
    \centering
    \begin{tabular}{|c|c|c|}
        \hline
         $\mathcal{L}_1$ & $\mathcal{L}_2$ & $\mathcal{L}_3$ \\
        \hline
         0.6208 & 0.9873 & 0.9908 \\
        \hline
    \end{tabular}
    \caption{Fidelity comparison between \name{}s trained with different loss functions.}
    \label{tab:acc_mnist_initial}
\end{table}
%\end{wraptable}
% \end{table}

Unfortunately, results in Table~\ref{tab:acc_mnist_initial} show that $\mathcal{L}_1$ achieves poor performance.
The average fidelity of prepared quantum states is only 0.6208.
As demonstrated in Fig.~\ref{fig:virtual_prep_states_losses},
$\mathcal{L}_1$ generates a state that losses the patterns of the original state.
% Unfortunately, training with $\mathcal{L}_1$ results in notable accuracy degradation as shown in Table~\ref{tab:acc_mnist_initial}.
Additionally, utilizing $\mathcal{L}_1$ implies that we need to first generate target parameters using AAE, of which the long runtime hinders pre-training on larger datasets.
Consequently, required is a more effective loss function design without involving AAE.

% \yilun{Below table needs fidelity comparison and mnist figure comparison.}

%\begin{wrapfigure}{r}{0.3\textwidth}
\begin{figure}[h]
    \centering
    \includegraphics[width=0.3\textwidth]{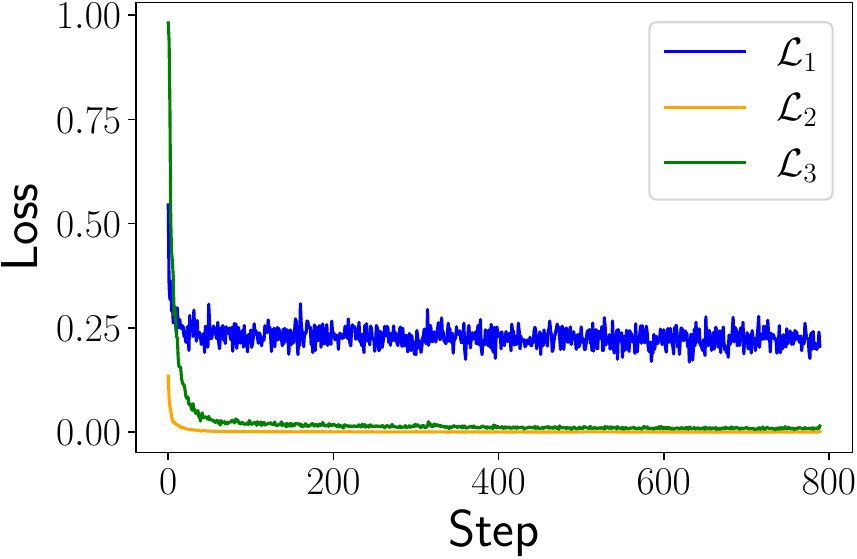}
    \caption{Convergence of different loss functions.}
    \label{fig:loss_comp}
\end{figure}
%\end{wrapfigure}

To address this challenge, we propose a \emph{state-oriented} training methodology, which employs quantum states as targets to guide optimizations.
% Specifically, by using quantum states as targets.
% Since our ultimate goal is to prepare a quantum state $\hat{\psi}$ that approximates the target state $\psi$,
Specifically,
we may apply $\boldsymbol{\hat{\theta}}$ to the circuit and execute it to obtain the prepared state $\hat{\psi}$.
Then it is possible to calculate the difference between $\hat{\psi}$ and $\psi$ as the loss to optimize \name{}.
In contrast to parameter-oriented training,
this approach applies to larger datasets
as it decouples the training procedure from AAE.
We utilize two different state-oriented metrics,
the first being the MSE between $\hat{\psi}$ and $\psi$,
denoted as $\mathcal{L}_2$,
and the second is the \emph{fidelity} of $\hat{\psi}$ relative to $\psi$, expressed as $\mathcal{L}_3 = 1 - |\langle\hat{\psi} |\psi\rangle|^2$~\cite{leung2017speedup_qoc}.
Results in Table~\ref{tab:acc_mnist_initial} show that $\mathcal{L}_2$ and $\mathcal{L}_3$ achieve remarkably higher fidelity than $\mathcal{L}_1$.
Besides, we observe that $\mathcal{L}_3$ prepares a state very similar to the target one (Fig.~\ref{fig:virtual_prep_states_losses}),
verifying that state-oriented training is more effective than parameter-oriented training.

\begin{figure}[h!]
    \begin{subfigure}{0.33\textwidth}
        \includegraphics[width=\linewidth]{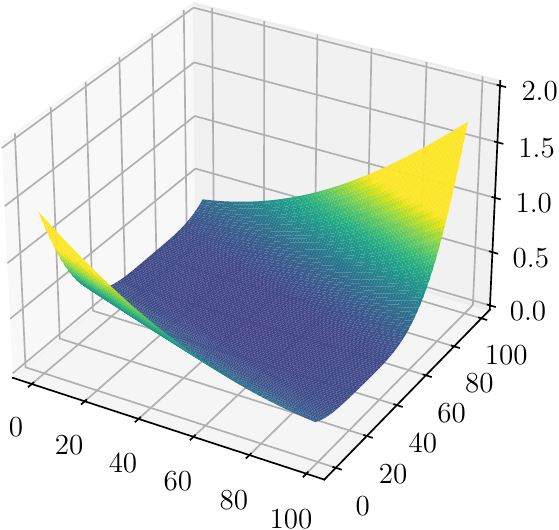}
        \caption{$\mathcal{L}_{1}$}
        \label{fig:landscape_mse}
    \end{subfigure}
    \begin{subfigure}{0.33\textwidth}
        \includegraphics[width=\linewidth]{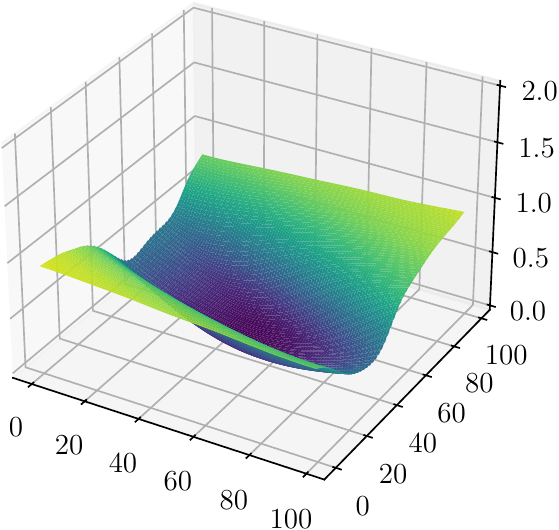}
        \caption{$\mathcal{L}_{2}$}
        \label{fig:landscape_state_mse}
    \end{subfigure}
    \begin{subfigure}{0.33\textwidth}
        \includegraphics[width=\linewidth]{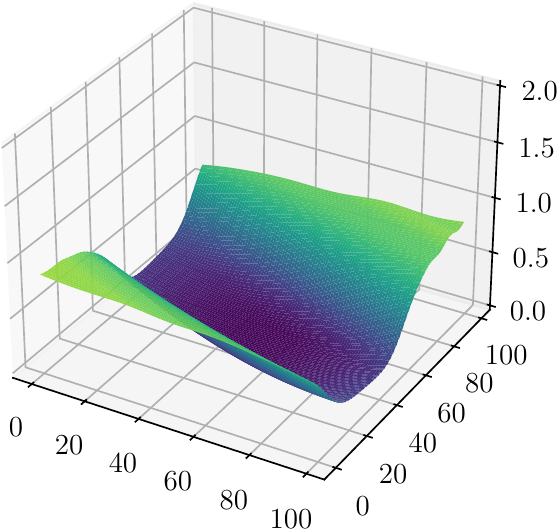}
        \caption{$\mathcal{L}_{3}$}
        \label{fig:landscape_state_fid}
    \end{subfigure}

    \caption{Landscape virtualization of different loss functions.}
    \label{fig:landscape}
\end{figure}

\noindent \textbf{Landscape Analysis}.
To understand the efficacy of these loss functions, we further analyze their landscapes following previous studies~\cite{li2018visualizing_loss_landscape,petzka2021relative_flatness_and_generalization_loss,christiancmehilwarn2022losslandscapes_blog}.
To gain insight from the landscape, we plot Fig.~\ref{fig:landscape} using the same scale and color gradients~\cite{christiancmehilwarn2022losslandscapes_blog}.
Compared to state-oriented losses ($\mathcal{L}_2$ and $\mathcal{L}_3$), $\mathcal{L}_1$ has a largely flat landscape with non-decreasing minima,
thus the model struggles to explore a viable path towards a lower loss value,
a similar pattern can also be observed in Fig.~\ref{fig:loss_comp}.
In contrast, $\mathcal{L}_2$ and $\mathcal{L}_3$ have much lower minima and successfully converge to smaller loss values.
Furthermore, we observe from Fig.~\ref{fig:landscape}
that $\mathcal{L}_3$ has a wider minima than $\mathcal{L}_2$, which may indicate a better generalization capability~\cite{petzka2021relative_flatness_and_generalization_loss}.
% From Fig.~\ref{fig:landscape}, we can also observe that $\mathcal{L}_3$ has a wider minima than $\mathcal{L}_2$, which may indicate a better generalization capability~\cite{petzka2021relative_flatness_and_generalization_loss}.

\noindent \textbf{Gradient Analysis}.
Based on the landscape analysis, we adopt $\mathcal{L}_3$ as the loss function to train \name{}.
% $\mathcal{L}_2$ relies on $\hat{\psi}$, a pure state that can only be acquired on ideal simulator, preventing its adoption on real quantum devices,
% where we only a mixed state $\hat{\rho}$ can be obtained.
% In contrast,
We note that $\mathcal{L}_3$ can be written as $1 - \langle\psi|\hat{\psi}\rangle\langle\hat{\psi}|\psi\rangle$.
If $\hat{\rho}$ is a pure state, it is equivalent to $|\hat{\psi}\rangle\langle\hat{\psi}|$.
% Thus we are able to use a general form of $\mathcal{L}_3$ as follows,
Then $\mathcal{L}_3$ is given by $\mathcal{L}_3 = 1 - \langle\psi | \hat{\rho} | \psi\rangle$.
% on both ideal simulation and noisy real devices.

% \begin{equation}
%     \label{eq:loss}
%     \mathcal{L}_3 = 1 - \langle\psi | \hat{\rho} | \psi\rangle.
% \end{equation}

% \todo{1. first gradient of L. 2. two cases of partial rho partial theta}
This re-formalization is important as only the mixed state $\hat{\rho}$ could be obtained in noisy environments.
Suppose an $n$-qubit circuit is parameterized by $m$ parameters $\boldsymbol{\hat{\theta}} = [\hat{\theta}_1, \dots, \hat{\theta}_k, \dots, \hat{\theta}_m]$.
Let $\mathbf{W}$ be the weight matrix of MLP, with $k,l$ the element indices.
We analyze the gradient of $\mathcal{L}_3$ w.r.t. $W_{k,l}$ to showcase its feasibility in different quantum computing environments.

\begin{equation}
    \begin{aligned}
        % \frac{\partial \mathcal{L}}{\partial  W_{k,l}} &=
        \nabla_{ W_{k,l}}\mathcal{L}_3
        &=
        \frac{\partial \mathcal{L}_3}{\partial W_{k,l}} =
        -\langle \psi|
        \frac{\partial\hat{\rho}}{\partial W_{k,l}}
        |\psi\rangle \\
        &=
        -\langle \psi|
        \begin{bmatrix}
            \sum_{j=1}^m
            \frac{\partial\hat{\rho}_{1,1}}{\partial\theta_{j}}
            % \cdot
            \frac{\partial\theta_j}{\partial W_{k,l}}
            & \cdots &
            \sum_{j=1}^m
            \frac{\partial\hat{\rho}_{1,N}}{\partial\theta_{j}}
            % \cdot
            \frac{\partial\theta_j}{\partial W_{k,l}} \\
            \vdots & \ddots & \vdots \\
            \sum_{j=1}^m
            \frac{\partial\hat{\rho}_{N,1}}{\partial\theta_{j}}
            % \cdot
            \frac{\partial\theta_j}{\partial W_{k,l}}
            & \cdots &
            \sum_{j=1}^m
            \frac{\partial\hat{\rho}_{N,N}}{\partial\theta_{j}}
            % \cdot
            \frac{\partial\theta_j}{\partial W_{k,l}} \\
        \end{bmatrix}
        % \left(
        % \frac{\partial\hat{\rho}_{i,j}}{\partial\boldsymbol{\hat{\theta}}}
        % \cdot
        % \frac{\partial\boldsymbol{\hat{\theta}}}{\partial W_{k,l}}
        % \right)_{i,j}
        |\psi\rangle,
    \end{aligned}
\end{equation}

The calculation of $\frac{\partial\theta_j}{\partial W_{k,l}}$ can be easily done on classical devices using backpropagation supported by automatic differentiation frameworks.
Therefore, we only focus on
$\frac{\partial\hat{\rho}_{i,j}}{\partial\theta_{k}}$.
In a simulation environment, the calculation of $\hat{\rho}$ is conducted via noisy quantum circuit simulation,
which is essentially a series of tensor operations on state vectors.
Therefore, the calculation of
$\frac{\partial\hat{\rho}_{i,j}}{\partial\theta_{k}}$
is compatible with backpropagation.
% However, backpropagation is not applicable on real quantum devices.
The situation on real devices becomes more complicated.
On real devices, the mixed state $\hat{\rho}$ is reconstructed through \emph{quantum tomography}~\cite{cramer2010efficient_tomo} based on classical shadow~\cite{zhang2021experimental_classical_shadow,huang2022learning_states_classical_shadow}.
Here, for notion simplicity, we denote the process of classical shadow as a transformation $\mathcal{S}$, and denote the measurement expectations of the ansatz as $U(\boldsymbol{\hat{\theta}})$.
Thus the reconstructed density matrix is given by $\hat{\rho} = \mathcal{S}(U(\boldsymbol{\hat{\theta}}))$.
Then the gradient of $\hat{\rho}_{i,j}$ with respect to $\hat{\theta}_{k}$ is
    % $\left(\frac{\partial\mathcal{S}(U(\boldsymbol{\hat{\theta}}))}{\partial U(\boldsymbol{\hat{\theta}})}\right)^T
    % \frac{\partial U(\boldsymbol{\hat{\theta}})}{\partial \theta_j}$.
$\sum_{u}
\frac{\partial\hat{\rho}_{i,j}}{\partial U(\boldsymbol{\hat{\theta}})}
    \frac{\partial U(\boldsymbol{\hat{\theta}})}{\partial \hat{\theta}_k}$.
Here $\frac{\partial\hat{\rho}_{i,j}}{\partial U(\boldsymbol{\hat{\theta}})}$
can be efficiently calculated on classical devices using backpropagation,
as $\mathcal{S}$ operates on expectation values on classical devices.
However, $U(\boldsymbol{\hat{\theta}})$ involves state evolution on quantum devices,
where back-propagation is impossible due to the No-Cloning theorem~\cite{nielsen2010qc-and-qi}.
% However, back-propagation on real quantum devices is infeasible due the No-Cloning theorem~
% \cite{nielsen2010qc-and-qi}.
Fortunately, it is possible to utilize the \emph{parameter shift} rule~\cite{crooks2019gradients_param_shift,bergholm2018pennylane_param_shift,wierichs2022general_param_shift} to calculate
$\frac{\partial U(\boldsymbol{\hat{\theta}})}{\partial \theta_k}$.
In this way, the gradients of the circuit function $U$ with respect to $\theta_j$ are
$\frac{\partial U(\boldsymbol{\hat{\theta}})}{\partial \theta_k} = \frac{1}{2}\left(
        U(\theta_{+}) - U(\theta_{-})
        \right)$,
where $\theta_{+} = [\theta_1,\dots,\theta_k+\frac{\pi}{2},\dots,\theta_m],\theta_{-} = [\theta_1,\dots,\theta_k-\frac{\pi}{2},\dots,\theta_m]$.
To summarize, training \name{} with $\mathcal{L}_3$ is theoretically feasible on both simulators and real devices.

% \begin{equation}
%     \begin{aligned}
%         \frac{\partial U(\boldsymbol{\hat{\theta}})}{\partial \theta_k}
%         &=
%         \frac{1}{2}\left(
%         U(\theta_{+}) - U(\theta_{-}),
%         \right) \\
%         \theta_{+} &= [\theta_1,\dots,\theta_k+\frac{\pi}{2},\dots,\theta_m], \\
%         \theta_{-} &= [\theta_1,\dots,\theta_k-\frac{\pi}{2},\dots,\theta_m].
%     \end{aligned}
% \end{equation}

% where $\frac{\partial\hat{\rho}}{\partial\boldsymbol{\hat{\theta}_{i,j}}}$ could be calculated via \emph{parameter shift}\todo{citation}.

\section{Numerical Results}

% In this section, we conduct a set of experiments to evaluate the performance of \name{} on both synthetic datasets and downstream tasks.
% Our primary objectives are to evaluate the runtime, scalability, and state fidelity against AE and AAE.
% Besides, we explore the robustness of \name{} in noisy environments.

\subsection{Experiment Setup}
\label{ssec:setup}

\noindent
\textbf{Datasets.}
To train a \name{} for arbitrary quantum states, we need a dataset comprising a wide range of quantum states with different distributions.
To our knowledge, there is no dataset dedicated for this special purpose.
A natural solution is to use readily available datasets from classical machine learning domains (e.g., ImageNet~\cite{deng2009imagenet}, Places~\cite{zhou2017places}, SQuAD~\cite{rajpurkar2016squad})
by normalizing them to quantum states.
However, QSP is essential in various application scenarios besides QML.
The classical data to be loaded may not only contain natural images or languages but also contain arbitrary data (e.g., in HHL algorithm~\cite{Harrow2009-ai_hhl}).
Therefore, we construct a training dataset adapted from FractalDB-60~\cite{kataoka2020fractal_db} with 60k samples,
a formula-driven dataset originally designed for computer vision without any natural images.
We also construct a separate dataset to test the performance of QSP, which consists of data sampled from different statistical distributions,
including uniform, normal, log-normal, exponential, and Dirichlet distributions,
with 3000 samples per distribution.
Hereafter we refer this dataset as the \emph{synthetic dataset}.

\noindent
\textbf{Platforms.}
We implement \name{} using PennyLane~\cite{mottonen2004pennylane_AME}, PyTorch~\cite{paszke2019pytorch} and Qiskit~\cite{Qiskit}.
Simulations are done on a Ubuntu server with 768 GB memory, two 32-core Intel(R) Xeon(R) Silver 4216 CPU with 2.10 GHz,
and 2 NVIDIA A-100 GPUs.
IBM quantum cloud platform\footnote{\url{https://quantum-computing.ibm.com/}} is adopted to evaluate the performance on real quantum devices.

\noindent
\textbf{Metrics.}
We evaluate \name{} and compare it to AE and AAE in terms of runtime, scalability, and fidelity.
\emph{Runtime} refers to how long it takes to prepare a quantum state.
\emph{Scalability} refers to how the circuit depth grows with the number of qubits.
\emph{Fidelity} evaluates the similarity between prepared quantum states and target quantum states. Specifically, the fidelity for two mixed states given by density matrices
$\rho$ and $\hat{\rho}$ is defined as $F(\rho,\hat{\rho}) = \mathrm{Tr}\left(\sqrt{\sqrt{\rho}\hat{\rho}\sqrt{\rho}}\right)^2 \in [0, 1]$.
A larger $F$ indicates a better fidelity.

\textbf{Implementation.}
We implement \name{} using an MLP consisting of two hidden layers.
The dimensions of input and output layers are respectively set to $2^n$ and $m$, where $n$ refers to the number of qubits and $m$ refers to the number of parameters.
We adopt $\mathcal{L}_3$ as the loss function.
Training data are down-sampled, flattened, and normalized to $2^n$-dimensional state vectors.
We adopt the hardware efficient ansatz~\cite{nature_hw_efficient_ansatz} (Fig.~\ref{fig:example_pqc}) as the backbone of quantum circuits and use the same structure for AAE.
Given a target state, a pre-trained \name{} model is invoked to generate parameters and thus the circuit for QSP.
While for AAE, we employ online iterations for each state.
For AE, the arithmetic decomposition method in PennyLane~\cite{mottonen2004pennylane_AME,bergholm2018pennylane_param_shift} is adopted.
We defer more details about implementation to Appendix~\ref{sec:impl_detail}.
%Our framework is open-source at \url{https://anonymous.4open.science/r/SuperEncoder-A733} with detailed instructions to reproduce our results.

\subsection{Evaluation on Synthetic Dataset}

\label{ssec:eval_synthetic_dataset}

For simplicity and without loss of generality,
% here we only discuss the results of 4-qubit QSP tasks.
we focus our discussion on the results of 4-qubit QSP tasks.
% Results of larger quantum states are discussed in Appendix~\ref{ssec:res_larger_qsp}.
The outcomes for larger quantum states are detailed in Appendix~\ref{ssec:res_larger_qsp}.
The parameters of both AAE and \name{} are optimized based on ideal quantum circuit simulation.

\textbf{Runtime.}
% The results of runtime and fidelity evaluated using synthetic dataset are listed in Table.~\ref{tab:comp_runtime_fidelity_ideal}.
The runtime and fidelity results, evaluated on the synthetic dataset, are presented in Table~\ref{tab:comp_runtime_fidelity_ideal}.
We observe that \name{} runs faster than AAE by orders of magnitudes and has a similar runtime to AE,
affirming that \name{} effectively overcomes the main drawback of AAE.

\begin{table}[ht]
    \centering
    \scalebox{0.75}{
    \begin{tabular}{|m{1.8cm}|m{1cm}|m{1.2cm}|m{1cm}|m{1.2cm}|m{1cm}|m{1.2cm}|}
        \hline
\rowcolor{lightgray}
        \multirow{2}{*}{} & \multicolumn{2}{c|}{AE} & \multicolumn{2}{c|}{AAE} & \multicolumn{2}{c|}{\name{}} \\ \cline{2-7}
\rowcolor{lightgray}
         & Fidelity & Runtime & Fidelity & Runtime & Fidelity & Runtime \\ \hline
        \centering Uniform     &       &       &    \centering 0.9996      &       &  \centering 0.9731   &  \arraybackslash  \\
        \centering Normal      &       &       &    \centering 0.9992      &       &  \centering 0.8201   &  \arraybackslash  \\
        \centering Log-normal  &       &       &    \centering 0.9993      &       &  \centering 0.9421   &  \arraybackslash  \\
        \centering Exponential &       &       &    \centering 0.9996      &       &  \centering 0.9464   &  \arraybackslash  \\
        \centering Dirichlet   &       &       &    \centering 0.9995      &       &  \centering 0.9737   &  \arraybackslash  \\ \hline
        \centering Average     & \centering 1.0000  &   \centering 0.0162 s    &   \centering 0.9994      &  \centering 5.0201 s          &  \centering 0.9310    &   \centering\arraybackslash 0.0397 s     \\ \hline
    \end{tabular}
    }
    \caption{Comparison between AE, AAE and \name{} in terms of runtime and fidelity.}
    \label{tab:comp_runtime_fidelity_ideal}
\end{table}

\begin{figure}
    \centering
    \begin{subfigure}{0.4\textwidth}
    \centering
    \includegraphics[width=0.8\linewidth]{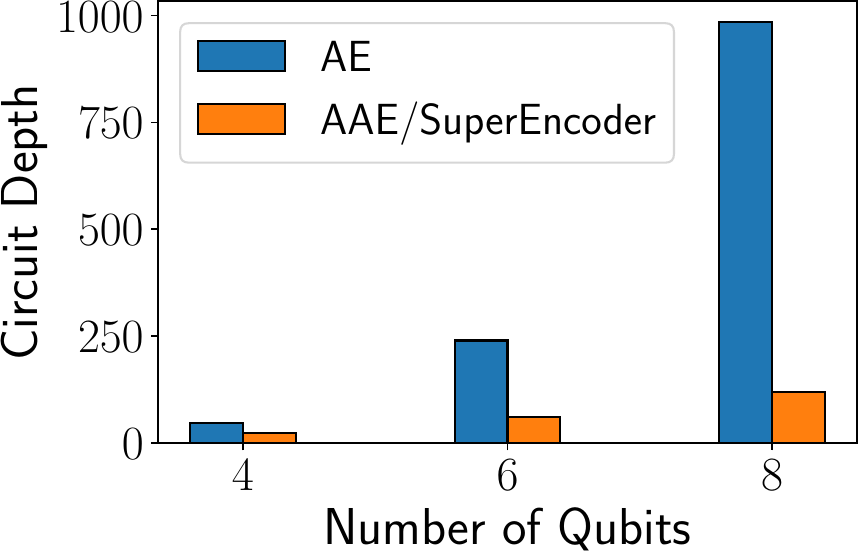}
        \caption{Scaling of circuit depth w.r.t. \# qubits.}
        \label{fig:scalability}
    \end{subfigure}
    \begin{subfigure}{0.5\textwidth}
        \centering
        \includegraphics[width=0.64\linewidth]{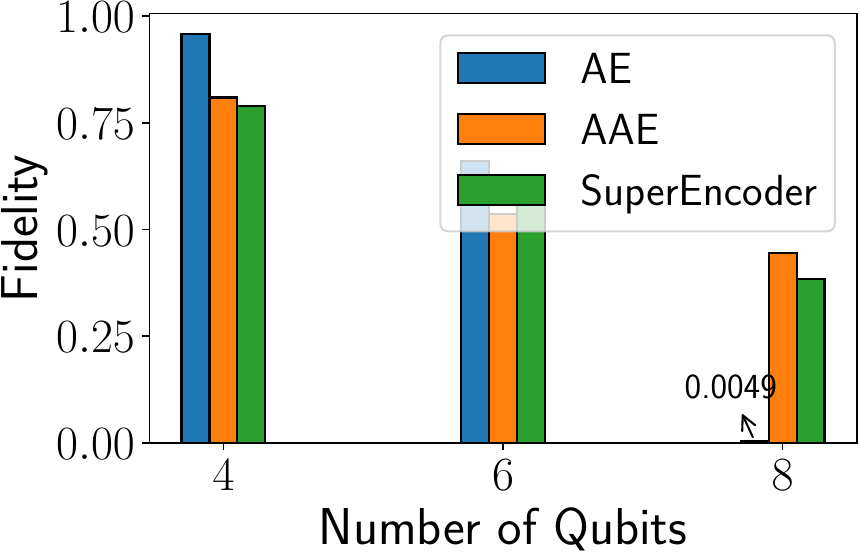}
        \caption{Fidelity of different QSP methods on \texttt{ibm\_osaka}.}
        \label{fig:fidelity_ae_real_device}
    \end{subfigure}
    \caption{Comparison between AE, AAE, and \name{} in terms of circuit depth and fidelity on real devices.}
    \label{fig:comp_three_methods_scalability}
\end{figure}

\noindent
\textbf{Scalability.}
Although AE runs fast, it exhibits poor scalability since the circuit depth grows exponentially with the number of qubits.
The depth of AAE is empirically determined by increasing depth until the final fidelity does not increase,
same depth is adopted for \name{}.
We deter the details of determining the depth of AAE/\name{} to Appendix~\ref{sec:impl_detail}.
As shown in Fig.~\ref{fig:scalability}, the depth of AE grows fast and becomes much larger than AAE/\name{},
e.g., the depth of AE for a 8-qubit state is 984, whereas the depth of AAE/\name{} is only 120.

% \begin{figure}
%     \centering
%     % \includegraphics[width=0.6\textwidth]{}
%     compare between AAE/\name{} and AE, 4, 6, 8, 10, 12 qubits
%     \caption{Scaling of circuit depth w.r.t \# qubits.}
%     \label{fig:scalability}
% \end{figure}

% \begin{figure}[h!]
%\begin{wrapfigure}{r}{0.4\textwidth}

\noindent
\textbf{Fidelity.}
% From Table.~\ref{tab:comp_runtime_fidelity_ideal}, we observe that \name{} has notable fidelity degradation compared with AAE and AE.
From Table~\ref{tab:comp_runtime_fidelity_ideal}, it is evident that \name{} experiences notable fidelity degradation when compared with AAE and AE.
% Specifically, the average fidelity of \name{} is 0.9307, while the average fidelity of AAE and AE is 0.9994 and 1.
Specifically, the average fidelity of \name{} is 0.9307, whereas AAE and AE achieve higher average fidelities of 0.9994 and 1.0, respectively.
Note that,
% although AE achieves best fidelity on ideal simulators, its effectiveness significantly degrades in noisy environments.
although AE demonstrates the highest fidelity under ideal simulation, its performance deteriorates significantly in noisy environments.
% As shown in Fig.~\ref{fig:fidelity_ae_real_device},
% the fidelity of AE for
Fig.~\ref{fig:fidelity_ae_real_device} presents the performance of these three QSP methods on quantum states with 4, 6, and 8 qubits on the \texttt{ibm\_osaka} machine.
While the fidelity of AE is higher than AAE/\name{} on the 4-qubit and 6-qubit states,
its fidelity on the 8-qubit state is only 0.0049,
becoming much lower than AAE/\name{}.
% We observe that the performance of AE significantly decreases on the 8-qubit state.
This decline is primarily attributed to its large circuit depth as shown in Fig.~\ref{fig:scalability}.

\subsection{Application to Downstream Tasks}

% \todo{ideal simualtion}

\begin{figure}[h]
    % \begin{subfigure}{0.4\textwidth}
    \centering
    \scalebox{0.5}{
    \begin{quantikz}
        & \gate[4,style={fill=red!20}]{\text{Encoder Block}}
        % \gategroup[wires=4,steps=3]{}
        & \gate[4]{U(\boldsymbol{\phi}_0)}
        % \gategroup[wires=4,steps=3,style={inner sep=6pt}]{entangler layers with tunable parameters}
        & \gate[4]{U(\boldsymbol{\phi}_1)} & \ \ldots \ & \gate[4]{U(\boldsymbol{\phi}_m)} & \qw\\
        & & & & \ \ldots \ & & \qw \\
        & & & & \ \ldots \ & & \qw \\
        & & & & \ \ldots \ & & \qw \\
    \end{quantikz}
    }
    % \vspace{-5pt}
    % \caption{Schematic of a QNN.}
    % \label{fig:schem_qnn}
    % \end{subfigure}
    % \begin{table}[ht!]
    % \begin{wraptable}{r}{0.4\textwidth}

    % \begin{subfigure}{0.4\textwidth}
        % \centering
        \scalebox{0.8}{
        \begin{tabular}{|c|c|c|}
            \hline
            AE & AAE & \name{} \\
            \hline
            97.15\% & 98.01\% & 97.87\% \\
            \hline
        \end{tabular}
        }
        % \caption{Test accuracies of different QSP methods on the QML task.}
        % \label{tab:acc_mnist_final}
    % \end{subfigure}
        % \vspace{-5pt}
    % \end{wraptable}
    % \end{table}
    % \label{fig:schem_qnn}
    % \vspace{-5pt}
% \end{wrapfigure}
    % \caption{}
% \end{figure}
    \caption{Schematic of a QNN (above) and test accuracies of QSP methods on the QML task (below).}
    \label{fig:schem_qnn}
%\end{wrapfigure}
\end{figure}
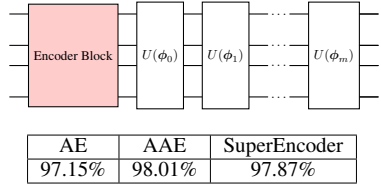

\textbf{Quantum Machine Learning.}
We first apply \name{} to a QML task.
MNIST dataset is adopted for demonstration, we extract a sub-dataset composed on digits 3 and 6 for evaluation.
% \vspace{-10pt}
% We construct the following quantum circuit (Fig.~\ref{fig:schem_qnn}) as a Quantum Neural Network (QNN),
The quantum circuit that implements a QNN is depicted in Fig.~\ref{fig:schem_qnn},
which consists of an encoder block and $m$ entangler layers.
Here the encoder block is implemented via QSP circuits, either AE, AAE, or \name{},
of which the parameters are frozen during the training of QNN.
The test results are shown in Fig.~\ref{fig:schem_qnn},
we observe that \name{} achieves similar performance with AAE and AE.
The reason lies in the fact that classification tasks can be robust to noises.
Consequently, approximate QSP (AAE and \name{}) with a certain degree of fidelity loss is tolerable.

\textbf{HHL Algorithm.}

%\begin{wrapfigure}{l}{0.3\textwidth}
\begin{figure}[h]
    \centering
    \includegraphics[width=0.3\textwidth, height=0.3\textwidth]{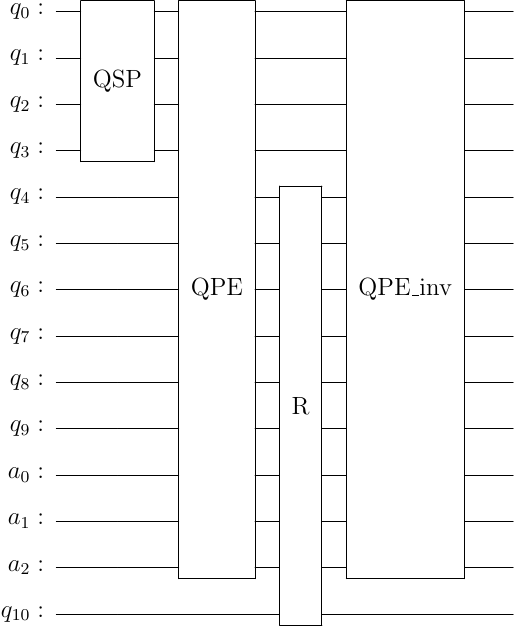}
    \caption{Schematic of HHL.}
    \label{fig:circ_hhl}
\end{figure}
%\end{wrapfigure}

Besides QML, quantum-enhanced linear algebra algorithms are another important set of applications that heavily rely on QSP.
The most famous algorithm is the HHL algorithm~\cite{Harrow2009-ai_hhl}.
The problem can be defined as,
given a matrix $\mathbf{A}\in\mathbb{C}^{N\times N}$, and a vector $\mathbf{b}\in\mathbb{C}^N$, find $\mathbf{x}\in\mathbb{C}^N$
satisfying $\mathbf{A}\mathbf{x} = \mathbf{b}$.
A typical implementation of HHL utilizes the circuit depicted in Fig.~\ref{fig:circ_hhl}.
The outline of HHL is as follows.
(i) Apply a QSP circuit to prepare the quantum state $|\mathbf{b}\rangle$.
(ii) Apply Quantum Phase Estimation~\cite{dorner2009optimal_qpe} (QPE) to estimate the eigenvalue of $\mathbf{A}$
(iii) Apply conditioned rotation gates on ancillary qubits based on the eigenvalues (R).
(iv) Apply an inverse QPE (QPE\_inv) and measure the ancillary qubits to reconstruct the solution vector $\mathbf{x}$.
Note that, HHL does not return the solution $\mathbf{x}$ itself,
but rather an approximation of the expectation value of some operator $\mathbf{M}$ associated with
$\mathbf{x}$, e.g., $\mathbf{x}^\dagger\mathbf{M}\mathbf{x}$.
Here, we adopt an optimized version of HHL proposed by Vazquez et al.~\cite{vazquez2022enhancing_hhl} for evaluation.
To compare the performance between different QSP methods,
we construct linear equations with fixed matrix $\mathbf{A}$ and operator $\mathbf{M}$,
while we sample different vectors from our synthetic dataset as $\mathbf{b}$.
% We test the performance of HHL by applying AE, AAE and \name{} as the backbone of its QSP circuit.
% The matrix $A$ is constructed by \todo{xxxx}
% We randomly sample 5 data from our synthetic dataset with different distributions to test the performance.
Results are concluded in Table~\ref{tab:res_hhl}.
Unlike QML, HHL expects precise QSP, thus we take the results from AE as the ground truth values and compare the relative error between AAE/\name{} and AE.
The relative error of \name{} is 2.4094\%, while the error of AAE is only 0.3326\%.

% \begin{figure}
%     \centering
%     \begin{subfigure}{0.3\textwidth}

%     \end{subfigure}
%     \begin{subfigure}

%     \end{subfigure}
%     \includegraphics{}
%     \caption{Caption}
%     \label{fig:enter-label}
% \end{figure}

\begin{table}[h]
%\begin{wraptable}{r}{0.4\textwidth}
    \centering
    \scalebox{0.75}{
    \begin{tabular}{|c|c|c|c|}
        \hline
\rowcolor{lightgray}
                        & AE & AAE & \name{} \\ \hline
        $\mathbf{b}_0$ & 0.7391 & 0.7404  & 0.7355  \\
        $\mathbf{b}_1$ & 0.7449 & 0.7445  & 0.7544   \\
        $\mathbf{b}_2$ & 0.7492 & 0.7469  & 0.8134   \\
        $\mathbf{b}_3$ & 0.7164 & 0.7099  & 0.7223   \\
        $\mathbf{b}_4$ & 0.7092 & 0.7076  & 0.7155   \\ \hline
        Avg err &  & 0.3326\% &  2.4094\%  \\ \hline
    \end{tabular}
    }
    \caption{Performance of different QSP methods in HHL algorithm. `Avg err' denotes the average relative errors between AAE/\name{} and AE.}
    \label{tab:res_hhl}
%\end{wraptable}
\end{table}

\subsection{Discussion and Future Work}
\label{ssec:disscuss_future_work}

The results of our evaluation can be concluded in two folds.
(i) \name{} effectively eliminates the iteration overhead of AAE, thereby becoming both fast and scalable.
However, it has a notable degradation in fidelity.
(ii) The impact of fidelity degradation varies across different downstream applications.
For QML, the fidelity degradation is affordable as long as the prepared states are distinguishable across different classes.
However, algorithms like HHL rely on precise QSP to produce the best result.
In these algorithms, \name{} suffers from higher error ratio than AAE.

Note that, the current evaluation results may not reflect the actual performance of \name{} on real NISQ devices.
Recent work has shown that AAE achieves significantly better fidelity than AE does~\cite{wang2023robuststate}.
This is due to the intrinsic noise awareness of AAE, as it could obtain states from noisy devices to guide updating parameters with better robustness.
In essence, the proposed \name{} possesses the same nature as AAE.
Unfortunately, although the noise-robustness of AAE can be evaluated on a small set of test samples,
it is difficult to perform noise-aware training for \name{} as it requires a large dataset for pre-training.
Consequently, \name{} relies on huge amounts of interactions with noisy devices, thereby becoming extremely time-consuming.
As a result, the effectiveness of \name{} in noisy environments remains largely unexplored, which we leave for future exploration.
More discussion about this perspective is in Appendix~\ref{sec:resource_est_real_device}.

\section{Related Work}

Besides QSP, there are other methods for loading classical data into quantum states.
These methods can be roughly regarded as \emph{quantum feature embedding} primarily used in QML,
which maps classical data to a completely different distribution encoded in quantum states.
A widely used embedding method is known as angle embedding.
Li et al. have proven that this method has a concentration issue, which means that the encoded states may become indistinguishable as the circuit depth increases~\cite{li2022concentration_AG}.
Lei et al. proposed an automatic design framework for efficient quantum feature embedding, resolving the issue of concentration~\cite{lei2023neural_auto_qemb}.
The central idea of this framework is to search for the most efficient circuit architecture for a given classical input, which is also known as Quantum Architecture Search  (QAS)~\cite{patel2024curriculum_qas,lu2023qas_bench,wu2023quantumdarts_qas}.
While the application scenario of quantum feature embedding is largely limited to QML, 
QSP has broader usage in general quantum applications, distinguishing \name{} from all aforementioned work.

\section{Conclusion}

In this work, we propose \name{}, a neural network-based QSP framework.
Instead of iteratively tuning the circuit parameters to approximate each quantum state, as is done in AAE,
we adopt a different approach by directly learning the relationship between target quantum states and the required circuit parameters.
\name{} combines the scalable circuit architecture of AAE with the fast runtime of AE, as verified by a comprehensive evaluation on both synthetic dataset and downstream applications.

\bibliographystyle{plain}
\bibliography{references}

\begin{thebibliography}{10}

\bibitem{abbas2021power_of_qnn}
Amira Abbas, David Sutter, Christa Zoufal, Aur{\'e}lien Lucchi, Alessio
  Figalli, and Stefan Woerner.
\newblock The power of quantum neural networks.
\newblock {\em Nature Computational Science}, 1(6):403--409, 2021.

\bibitem{araujo2023configurable_sub_qip_qsp}
Israel~F Araujo, Daniel~K Park, Teresa~B Ludermir, Wilson~R Oliveira, Francesco
  Petruccione, and Adenilton~J Da~Silva.
\newblock Configurable sublinear circuits for quantum state preparation.
\newblock {\em Quantum Information Processing}, 22(2):123, 2023.

\bibitem{bausch2020recurrent_qnn}
Johannes Bausch.
\newblock Recurrent quantum neural networks.
\newblock {\em Advances in neural information processing systems},
  33:1368--1379, 2020.

\bibitem{bergholm2018pennylane_param_shift}
Ville Bergholm, Josh Izaac, Maria Schuld, Christian Gogolin, Shahnawaz Ahmed,
  Vishnu Ajith, M~Sohaib Alam, Guillermo Alonso-Linaje, B~AkashNarayanan, Ali
  Asadi, et~al.
\newblock Pennylane: Automatic differentiation of hybrid quantum-classical
  computations.
\newblock {\em arXiv preprint arXiv:1811.04968}, 2018.

\bibitem{biamonte2017qml_nature_review}
Jacob Biamonte, Peter Wittek, Nicola Pancotti, Patrick Rebentrost, Nathan
  Wiebe, and Seth Lloyd.
\newblock Quantum machine learning.
\newblock {\em Nature}, 549(7671):195--202, 2017.

\bibitem{operatorapproximation_op_learn}
Tianping Chen and Hong Chen.
\newblock Universal approximation to nonlinear operators by neural networks
  with arbitrary activation functions and its application to dynamical systems.
\newblock {\em IEEE Transactions on Neural Networks}, 6(4):911--917, 1995.

\bibitem{cramer2010efficient_tomo}
Marcus Cramer, Martin~B Plenio, Steven~T Flammia, Rolando Somma, David Gross,
  Stephen~D Bartlett, Olivier Landon-Cardinal, David Poulin, and Yi-Kai Liu.
\newblock Efficient quantum state tomography.
\newblock {\em Nature communications}, 1(1):149, 2010.

\bibitem{crooks2019gradients_param_shift}
Gavin~E Crooks.
\newblock Gradients of parameterized quantum gates using the parameter-shift
  rule and gate decomposition.
\newblock {\em arXiv preprint arXiv:1905.13311}, 2019.

\bibitem{deng2009imagenet}
Jia Deng, Wei Dong, Richard Socher, Li-Jia Li, Kai Li, and Li~Fei-Fei.
\newblock Imagenet: A large-scale hierarchical image database.
\newblock In {\em 2009 IEEE conference on computer vision and pattern
  recognition}, pages 248--255. Ieee, 2009.

\bibitem{dorner2009optimal_qpe}
Uwe Dorner, Rafal Demkowicz-Dobrzanski, Brian~J Smith, Jeff~S Lundeen, Wojciech
  Wasilewski, Konrad Banaszek, and Ian~A Walmsley.
\newblock Optimal quantum phase estimation.
\newblock {\em Physical review letters}, 102(4):040403, 2009.

\bibitem{farhi2014quantum_qaoa}
Edward Farhi, Jeffrey Goldstone, and Sam Gutmann.
\newblock A quantum approximate optimization algorithm.
\newblock {\em arXiv preprint arXiv:1411.4028}, 2014.
\newblock \url{https://doi.org/10.48550/arXiv.1411.4028}.

\bibitem{gleinig2021efficient_special_sparse_qsp}
Niels Gleinig and Torsten Hoefler.
\newblock An efficient algorithm for sparse quantum state preparation.
\newblock In {\em 2021 58th ACM/IEEE Design Automation Conference (DAC)}, pages
  433--438. IEEE, 2021.

\bibitem{gonzalez2021simulate_option_price_PDE}
Javier Gonzalez-Conde, {\'A}ngel Rodr{\'\i}guez-Rozas, Enrique Solano, and
  Mikel Sanz.
\newblock Simulating option price dynamics with exponential quantum speedup.
\newblock {\em arXiv preprint arXiv:2101.04023}, 2021.

\bibitem{gonzalez2024efficient_special_qsp}
Javier Gonzalez-Conde, Thomas~W Watts, Pablo Rodriguez-Grasa, and Mikel Sanz.
\newblock Efficient quantum amplitude encoding of polynomial functions.
\newblock {\em Quantum}, 8:1297, 2024.

\bibitem{Harrow2009-ai_hhl}
Aram~W Harrow, Avinatan Hassidim, and Seth Lloyd.
\newblock Quantum algorithm for linear systems of equations.
\newblock {\em Physical Review Letters}, 103(15):150502, 2009.
\newblock \url{https://doi.org/10.1103/PhysRevLett.103.150502}.

\bibitem{huang2022learning_states_classical_shadow}
Hsin-Yuan Huang.
\newblock Learning quantum states from their classical shadows.
\newblock {\em Nature Reviews Physics}, 4(2):81--81, 2022.

\bibitem{iaconis2024qsp_normal_distribution}
Jason Iaconis, Sonika Johri, and Elton~Yechao Zhu.
\newblock Quantum state preparation of normal distributions using matrix
  product states.
\newblock {\em npj Quantum Information}, 10(1):15, 2024.

\bibitem{christiancmehilwarn2022losslandscapes_blog}
Christian Cmehil-Warn Jacob~Hansen.
\newblock Loss landscapes.
\newblock In {\em ICLR Blog Track}, 2022.
\newblock
  https://loss-landscapes.github.io/Loss-Landscapes-Blog/2022/12/01/loss-landscapes/.

\bibitem{jiang2021quantumflow_ncomm}
Weiwen Jiang, Jinjun Xiong, and Yiyu Shi.
\newblock A co-design framework of neural networks and quantum circuits towards
  quantum advantage.
\newblock {\em Nature Communications}, 12(1):579, 2021.
\newblock \url{https://doi.org/10.1038/s41467-020-20729-5}.

\bibitem{nature_hw_efficient_ansatz}
Abhinav Kandala, Antonio Mezzacapo, Kristan Temme, Maika Takita, Markus Brink,
  Jerry~M. Chow, and Jay~M. Gambetta.
\newblock Hardware-efficient variational quantum eigensolver for small
  molecules and quantum magnets.
\newblock {\em Nature}, 549(7671):242--246, September 2017.

\bibitem{kataoka2020fractal_db}
Hirokatsu Kataoka, Kazushige Okayasu, Asato Matsumoto, Eisuke Yamagata, Ryosuke
  Yamada, Nakamasa Inoue, Akio Nakamura, and Yutaka Satoh.
\newblock Pre-training without natural images.
\newblock In {\em Proceedings of the Asian Conference on Computer Vision},
  2020.

\bibitem{kingma2014adam}
Diederik~P Kingma and Jimmy Ba.
\newblock Adam: A method for stochastic optimization.
\newblock {\em arXiv preprint arXiv:1412.6980}, 2014.

\bibitem{krantz2019quantum_eng_guide_sc}
Philip Krantz, Morten Kjaergaard, Fei Yan, Terry~P Orlando, Simon Gustavsson,
  and William~D Oliver.
\newblock A quantum engineer's guide to superconducting qubits.
\newblock {\em Applied physics reviews}, 6(2), 2019.

\bibitem{lei2023neural_auto_qemb}
Cong Lei, Yuxuan Du, Peng Mi, Jun Yu, and Tongliang Liu.
\newblock Neural auto-designer for enhanced quantum kernels.
\newblock In {\em The Twelfth International Conference on Learning
  Representations}, 2023.

\bibitem{leung2017speedup_qoc}
Nelson Leung, Mohamed Abdelhafez, Jens Koch, and David Schuster.
\newblock Speedup for quantum optimal control from automatic differentiation
  based on graphics processing units.
\newblock {\em Physical Review A}, 95(4):042318, 2017.
\newblock \url{https://doi.org/10.1103/PhysRevA.95.042318}.

\bibitem{li2022concentration_AG}
Guangxi Li, Ruilin Ye, Xuanqiang Zhao, and Xin Wang.
\newblock Concentration of data encoding in parameterized quantum circuits.
\newblock {\em Advances in Neural Information Processing Systems},
  35:19456--19469, 2022.

\bibitem{li2019tackling}
Gushu Li, Yufei Ding, and Yuan Xie.
\newblock Tackling the qubit mapping problem for nisq-era quantum devices.
\newblock In {\em Proceedings of the Twenty-Fourth International Conference on
  Architectural Support for Programming Languages and Operating Systems}, pages
  1001--1014, 2019.
\newblock \url{https://doi.org/10.1145/3297858.3304023}.

\bibitem{li2018visualizing_loss_landscape}
Hao Li, Zheng Xu, Gavin Taylor, Christoph Studer, and Tom Goldstein.
\newblock Visualizing the loss landscape of neural nets.
\newblock {\em Advances in neural information processing systems}, 31, 2018.

\bibitem{long2001efficient_qsp}
Gui-Lu Long and Yang Sun.
\newblock Efficient scheme for initializing a quantum register with an
  arbitrary superposed state.
\newblock {\em Physical Review A}, 64(1):014303, 2001.

\bibitem{lu2023qas_bench}
Xudong Lu, Kaisen Pan, Ge~Yan, Jiaming Shan, Wenjie Wu, and Junchi Yan.
\newblock Qas-bench: rethinking quantum architecture search and a benchmark.
\newblock In {\em International Conference on Machine Learning}, pages
  22880--22898. PMLR, 2023.

\bibitem{lubasch2020vqa_for_nonlinear_problems}
Michael Lubasch, Jaewoo Joo, Pierre Moinier, Martin Kiffner, and Dieter Jaksch.
\newblock Variational quantum algorithms for nonlinear problems.
\newblock {\em Physical Review A}, 101(1):010301, 2020.

\bibitem{mao2024towards_sparse_special_qsp}
Rui Mao, Guojing Tian, and Xiaoming Sun.
\newblock Towards optimal circuit size for sparse quantum state preparation.
\newblock {\em arXiv e-prints}, pages arXiv--2404, 2024.

\bibitem{mitarai2018quantum_circuit_learning}
Kosuke Mitarai, Makoto Negoro, Masahiro Kitagawa, and Keisuke Fujii.
\newblock Quantum circuit learning.
\newblock {\em Physical Review A}, 98(3):032309, 2018.

\bibitem{mottonen2004pennylane_AME}
Mikko M{\"o}tt{\"o}nen, JJ~Vartiainen, Ville Bergholm, and Martti~M Salomaa.
\newblock Transformation of quantum states using uniformly controlled
  rotations.
\newblock {\em Quantum Information and Computation}, 5, 2005.

\bibitem{nakaji2022aae_pr_research}
Kouhei Nakaji, Shumpei Uno, Yohichi Suzuki, Rudy Raymond, Tamiya Onodera,
  Tomoki Tanaka, Hiroyuki Tezuka, Naoki Mitsuda, and Naoki Yamamoto.
\newblock Approximate amplitude encoding in shallow parameterized quantum
  circuits and its application to financial market indicators.
\newblock {\em Physical Review Research}, 4(2):023136, 2022.

\bibitem{nielsen2010qc-and-qi}
Michael~A Nielsen and Isaac~L Chuang.
\newblock Quantum computation and quantum information.
\newblock 2010.

\bibitem{paszke2019pytorch}
Adam Paszke, Sam Gross, Francisco Massa, Adam Lerer, James Bradbury, Gregory
  Chanan, Trevor Killeen, Zeming Lin, Natalia Gimelshein, Luca Antiga, et~al.
\newblock Pytorch: An imperative style, high-performance deep learning library.
\newblock {\em Advances in neural information processing systems}, 32, 2019.

\bibitem{patel2024curriculum_qas}
Yash~J. Patel, Akash Kundu, Mateusz Ostaszewski, Xavier Bonet-Monroig, Vedran
  Dunjko, and Onur Danaci.
\newblock Curriculum reinforcement learning for quantum architecture search
  under hardware errors.
\newblock In {\em The Twelfth International Conference on Learning
  Representations}, 2024.

\bibitem{vqe_nature_comm}
Alberto Peruzzo, Jarrod McClean, Peter Shadbolt, Man-Hong Yung, Xiao-Qi Zhou,
  Peter~J Love, Al{\'a}n Aspuru-Guzik, and Jeremy~L O’brien.
\newblock A variational eigenvalue solver on a photonic quantum processor.
\newblock {\em Nature communications}, 5(1):4213, 2014.
\newblock \url{https://doi.org/10.1038/ncomms5213}.

\bibitem{petzka2021relative_flatness_and_generalization_loss}
Henning Petzka, Michael Kamp, Linara Adilova, Cristian Sminchisescu, and Mario
  Boley.
\newblock Relative flatness and generalization.
\newblock {\em Advances in neural information processing systems},
  34:18420--18432, 2021.

\bibitem{plesch2011qsp_PRA}
Martin Plesch and {\v{C}}aslav Brukner.
\newblock Quantum-state preparation with universal gate decompositions.
\newblock {\em Physical Review A}, 83(3):032302, 2011.

\bibitem{preskill2018quantum}
John Preskill.
\newblock Quantum computing in the {NISQ} era and beyond.
\newblock {\em Quantum}, 2:79, 2018.

\bibitem{Qiskit}
{Qiskit contributors}.
\newblock Qiskit: An open-source framework for quantum computing, 2023.

\bibitem{rajpurkar2016squad}
Pranav Rajpurkar, Jian Zhang, Konstantin Lopyrev, and Percy Liang.
\newblock Squad: 100,000+ questions for machine comprehension of text.
\newblock {\em arXiv preprint arXiv:1606.05250}, 2016.

\bibitem{schuld2016prediction_linear_regress_algebra}
Maria Schuld, Ilya Sinayskiy, and Francesco Petruccione.
\newblock Prediction by linear regression on a quantum computer.
\newblock {\em Physical Review A}, 94(2):022342, 2016.

\bibitem{shende2005synthesis_qsp}
Vivek~V Shende, Stephen~S Bullock, and Igor~L Markov.
\newblock Synthesis of quantum logic circuits.
\newblock In {\em Proceedings of the 2005 Asia and South Pacific Design
  Automation Conference}, pages 272--275, 2005.

\bibitem{shor1999polynomial}
Peter~W Shor.
\newblock Polynomial-time algorithms for prime factorization and discrete
  logarithms on a quantum computer.
\newblock {\em SIAM review}, 41(2):303--332, 1999.
\newblock \url{https://doi.org/10.1137/S0036144598347011}.

\bibitem{srinivasan2018learning_infer_hilbert_nips}
Siddarth Srinivasan, Carlton Downey, and Byron Boots.
\newblock Learning and inference in hilbert space with quantum graphical
  models.
\newblock {\em Advances in Neural Information Processing Systems}, 31, 2018.

\bibitem{sun2023asymptotically_optimal_qsp}
Xiaoming Sun, Guojing Tian, Shuai Yang, Pei Yuan, and Shengyu Zhang.
\newblock Asymptotically optimal circuit depth for quantum state preparation
  and general unitary synthesis.
\newblock {\em IEEE Transactions on Computer-Aided Design of Integrated
  Circuits and Systems}, 2023.

\bibitem{tian2023recent_qnn_survey}
Jinkai Tian, Xiaoyu Sun, Yuxuan Du, Shanshan Zhao, Qing Liu, Kaining Zhang, Wei
  Yi, Wanrong Huang, Chaoyue Wang, Xingyao Wu, et~al.
\newblock Recent advances for quantum neural networks in generative learning.
\newblock {\em IEEE Transactions on Pattern Analysis and Machine Intelligence},
  2023.

\bibitem{vazquez2022enhancing_hhl}
Almudena~Carrera Vazquez, Ralf Hiptmair, and Stefan Woerner.
\newblock Enhancing the quantum linear systems algorithm using richardson
  extrapolation.
\newblock {\em ACM Transactions on Quantum Computing}, 3(1):1--37, 2022.

\bibitem{wang2023robuststate}
Hanrui Wang, Yilian Liu, Pengyu Liu, Jiaqi Gu, Zirui Li, Zhiding Liang, Jinglei
  Cheng, Yongshan Ding, Xuehai Qian, Yiyu Shi, et~al.
\newblock Robuststate: Boosting fidelity of quantum state preparation via
  noise-aware variational training.
\newblock {\em arXiv preprint arXiv:2311.16035}, 2023.

\bibitem{wierichs2022general_param_shift}
David Wierichs, Josh Izaac, Cody Wang, and Cedric Yen-Yu Lin.
\newblock General parameter-shift rules for quantum gradients.
\newblock {\em Quantum}, 6:677, 2022.

\bibitem{wu2023quantumdarts_qas}
Wenjie Wu, Ge~Yan, Xudong Lu, Kaisen Pan, and Junchi Yan.
\newblock Quantumdarts: differentiable quantum architecture search for
  variational quantum algorithms.
\newblock In {\em International Conference on Machine Learning}, pages
  37745--37764. PMLR, 2023.

\bibitem{zhang2021experimental_classical_shadow}
Ting Zhang, Jinzhao Sun, Xiao-Xu Fang, Xiao-Ming Zhang, Xiao Yuan, and He~Lu.
\newblock Experimental quantum state measurement with classical shadows.
\newblock {\em Physical Review Letters}, 127(20):200501, 2021.

\bibitem{zhang2021low_depth_QSP_ancilla}
Xiao-Ming Zhang, Man-Hong Yung, and Xiao Yuan.
\newblock Low-depth quantum state preparation.
\newblock {\em Physical Review Research}, 3(4):043200, 2021.

\bibitem{zhao2019state_ancilla_qpe}
Jian Zhao, Yu-Chun Wu, Guang-Can Guo, and Guo-Ping Guo.
\newblock State preparation based on quantum phase estimation.
\newblock {\em arXiv preprint arXiv:1912.05335}, 2019.

\bibitem{zhou2017places}
Bolei Zhou, Agata Lapedriza, Aditya Khosla, Aude Oliva, and Antonio Torralba.
\newblock Places: A 10 million image database for scene recognition.
\newblock {\em IEEE transactions on pattern analysis and machine intelligence},
  40(6):1452--1464, 2017.

\bibitem{zoufal2019quantum_gan_npjq}
Christa Zoufal, Aur{\'e}lien Lucchi, and Stefan Woerner.
\newblock Quantum generative adversarial networks for learning and loading
  random distributions.
\newblock {\em npj Quantum Information}, 5(1):103, 2019.

\end{thebibliography}

% {
% \small

% [1] Alexander, J.A.\ \& Mozer, M.C.\ (1995) Template-based algorithms for
% connectionist rule extraction. In G.\ Tesauro, D.S.\ Touretzky and T.K.\ Leen
% (eds.), {\it Advances in Neural Information Processing Systems 7},
% pp.\ 609--616. Cambridge, MA: MIT Press.

% [2] Bower, J.M.\ \& Beeman, D.\ (1995) {\it The Book of GENESIS: Exploring
%   Realistic Neural Models with the GEneral NEural SImulation System.}  New York:
% TELOS/Springer--Verlag.

% [3] Hasselmo, M.E., Schnell, E.\ \& Barkai, E.\ (1995) Dynamics of learning and
% recall at excitatory recurrent synapses and cholinergic modulation in rat
% hippocampal region CA3. {\it Journal of Neuroscience} {\bf 15}(7):5249-5262.
% }

%%%%%%%%%%%%%%%%%%%%%%%%%%%%%%%%%%%%%%%%%%%%%%%%%%%%%%%%%%%%

%%%%%%%%%%%%%%%%%%%%%%%%%%%%%%%%%%%%%%%%%%%%%%%%%%%%%%%%%%%%

\newpage

\appendix

The structure of our Appendix is as follows. 
Appendix~\ref{sec:impl_detail} provides more details of implementing \name{}.
Appendix~\ref{sec:more_res} provides additional numerical results to illustrate the impact of state sizes, model architectures, and training datasets.
Appendix~\ref{sec:resource_est_real_device} analyzes the estimated runtime of training \name{} on real devices.
% Additionally, we demonstrate the feasibility of training \name{} in a noisy environment.

% \todo{parameter shift; classical shadow ($\mathcal{S}$)}

\section{Implementation Details}

\label{sec:impl_detail}

In this section, we elaborate the missing details of \name{} in the main text.

\begin{figure}[h]
    \centering
    \includegraphics{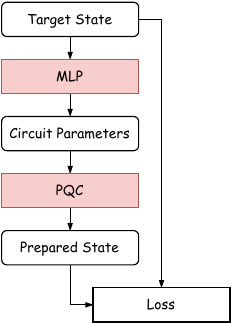}
    \caption{Detailed workflow of \name{}.}
    \label{fig:se_detail_workflow}
\end{figure}

The overarching workflow of \name{} is illustrated in Fig.~\ref{fig:se_detail_workflow}.
The target quantum states are input to the MLP model.
Then, the MLP model generates predicted parameters based on the target states.
Afterwards, the parameters are applied to the PQC to obtain the prepared quantum states.
Finally, we calculate the loss based on the prepared states and target states and optimize the weights of MLP through backpropagation.

The settings of MLP and PQC are as follows.

\noindent
\textbf{MLP.}
As listed in Table~\ref{tab:mlp_config}, we implement a two-layer MLP.
Each layer consists of 512 neurons.
We employ \texttt{Tanh} as the activation functions since $\boldsymbol{\theta}$ represents the \emph{angles} of rotation gates, ranging from $-\pi$ to $\pi$.

\begin{table}[!h]
        \centering
        
\begin{tabular}{|p{0.2\textwidth}|p{0.2\textwidth}|p{0.25\textwidth}|}
\hline 
 \multirow{2}{*}{Linear} & $\displaystyle \text{Input}$ & $\displaystyle \left(\text{batch\_size} ,\ 2^{n}\right)$ \\
\cline{2-3} 
   & $\displaystyle \text{Output}$ & $\displaystyle \left(\text{batch\_size} ,\ 512\right)$ \\
\hline 
 \multirow{2}{*}{Tanh} & $\displaystyle \text{Input}$ & $\displaystyle \left(\text{batch\_size} ,\ 512\right)$ \\
\cline{2-3} 
   & $\displaystyle \text{Output}$ & $\displaystyle \left(\text{batch\_size} ,\ 512\right)$ \\
\hline 
 \multirow{2}{*}{Linear} & $\displaystyle \text{Input}$ & $\displaystyle \left(\text{batch\_size} ,\ 512\right)$ \\
\cline{2-3} 
   & $\displaystyle \text{Output}$ & $\displaystyle \left(\text{batch\_size} ,\ \dim(\boldsymbol{\theta })\right)$ \\
\hline 
 \multirow{2}{*}{Tanh} & $\displaystyle \text{Input}$ & $\displaystyle \left(\text{batch\_size} ,\ \dim(\boldsymbol{\theta })\right)$ \\
\cline{2-3} 
   & $\displaystyle \text{Output}$ & $\displaystyle \left(\text{batch\_size} ,\ \dim(\boldsymbol{\theta })\right)$ \\
 \hline
\end{tabular}

\caption{MLP based \name{}. $n$ refers to the number of qubits. $\boldsymbol{\theta}$ denotes the parameter vector.}
\label{tab:mlp_config}
        
\end{table}

\begin{figure}[h]
    \centering
    \includegraphics[width=0.6\textwidth]{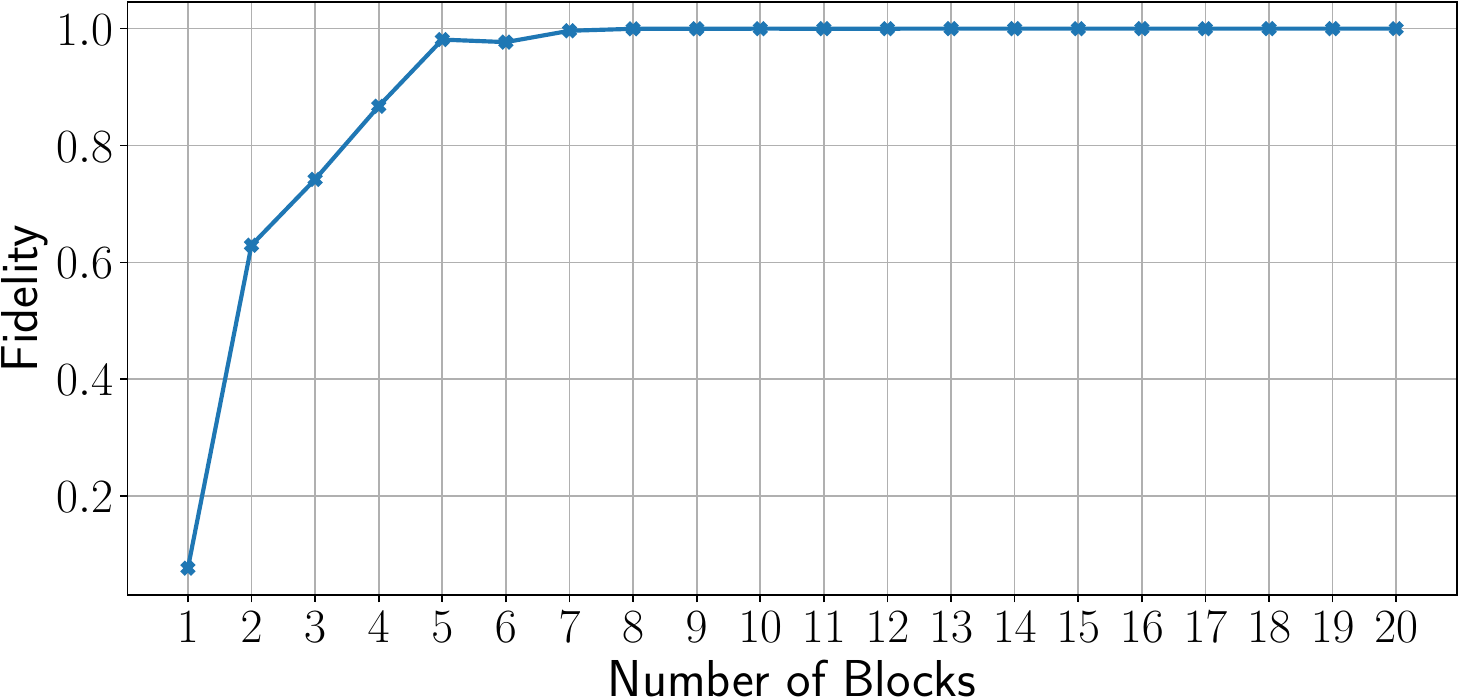}
    \caption{Fidelity vs. \# blocks for 4-qubit states using AAE.}
    \label{fig:fid_depth_aae}
\end{figure}

\textbf{PQC.}
The circuit structure is the same with the one depicted in Fig.~\ref{fig:example_pqc},
except that the number of blocks is determined dynamically through empirical examinations.
Specifically, we utilize AAE to approximate a target state while increasing the number of blocks.
The number of blocks is designated when the resulting state fidelity no longer increases.
For example, Fig.~\ref{fig:fid_depth_aae} demonstrates how fidelity changes while increasing the number of blocks.
As one can observe, the fidelity converges when the number of layers is larger than 8. 
Hence, the number of layers is set to be 8 for 4-qubit quantum states.
We follow the same procedure to set the number of blocks for other state sizes.
Each block has the same structure, consisting of a rotation layer and an entangler layer.
Given an $n$-qubit system, a rotation layer comprises $n$ $R_y$ gates, each operating on a distinct qubit.
The entangler layer is composed of two CNOT layers.
The first CNOT layer applies CNOT gates to $\{(q_0,q_1),(q_2,q_3),\dots\}$, and the second CNOT layer applies CNOT gates to $\{(q_1,q_2),(q_3,q_4),\dots\}$.
Hence, the depth of a block is 3.
Let $l$ be the number of blocks; then the dimension of the parameter vector is given by $\dim(\boldsymbol{\theta}) = n \times l$,
and the depth of AAE/\name{} is $3\times l$.
We conclude the settings of AAE/\name{} used throughout this study in Table~\ref{tab:aae_settings}.

\begin{table}[!h]
        \centering
        
\begin{tabular}{|p{0.2\textwidth}|p{0.05\textwidth}|p{0.05\textwidth}|p{0.05\textwidth}|}
\hline 
 Number of Qubits & 4 & 6 & 8 \\
\hline 
 Number of Blocks & 8 & 20 & 40 \\
\hline 
 Depth & 24 & 60 & 120 \\
 \hline
\end{tabular}

\caption{Number of blocks and corresponding depth of AAE/\name{}.}
\label{tab:aae_settings}  
\end{table}

The hyperparameters for training \name{} and optimizing AAE are as follows.

\noindent
\textbf{Training Hyperparameters for \name{}.}
Throughout our experiments, the number of epochs are consistently set to be 10.
For 4-qubit states, we set \texttt{bath\_size} to 32, while we set it 64 for 6-qubit and 8-qubit states.
We adopt Adam optimizer~\cite{kingma2014adam} with a learning rate of 3e-3 and a weight decay of 1e-5.

\noindent
\textbf{Hyperparameters for AAE.}
To optimize the parameters of AAE, we also use the Adam optimizer, with a learning rate of 1e-2 and zero weight decay.
For all quantum states, we train the AAE for 100 steps.

% \section{Appendix /supplemental material}
\section{More Numerical Results}

\label{sec:more_res}

\subsection{Results on Larger Quantum States}

\label{ssec:res_larger_qsp}

% Since the size of quantum states grows exponentially with the number of qubits,
% training a \name{} quickly becomes extremely slow.
% However, this issue can be alleviated by the 

In line with the main text, we train the \name{} for 6-qubit and 8-qubit quantum states using FractalDB-60 as the training dataset.
Then we evaluate the performance of \name{} on the synthetic test datasets.
As shown in Table~\ref{tab:larger_states}, the average fidelity on 6-qubit and 8-qubit states are 0.8655 and 0.7624 respectively.
In Appendix~\ref{ssec:impact_model_arch},~\ref{ssec:impact_training_dataset},
we discuss potential optimizations to alleviate this performance degradation.

\begin{table}[!h]
        \centering
        
\begin{tabular}{|p{0.15\textwidth}|p{0.1\textwidth}|p{0.1\textwidth}|p{0.1\textwidth}|}
\hline 
\rowcolor{lightgray}
% \cellcolor{gray} 
\centering Dataset & \centering$\displaystyle n=4$ & \centering$\displaystyle n=6$ & \centering\arraybackslash$\displaystyle n=8$ \\
\hline 
 \centering Uniform & \centering 0.9731 & \centering 0.9254 & \centering\arraybackslash 0.8648 \\
\hline 
 \centering Normal & \centering 0.8201 & \centering 0.7457 & \centering\arraybackslash 0.6075 \\
\hline 
 \centering Log-normal & \centering 0.9421 & \centering 0.8575 & \centering\arraybackslash 0.7122 \\
\hline 
 \centering Exponential & \centering 0.9464 & \centering 0.8757 & \centering\arraybackslash 0.7613 \\
\hline 
  \centering Dirichlet & \centering 0.9737 & \centering 0.9232 & \centering\arraybackslash 0.8663 \\
\hline  \centering
 Avg & \centering 0.9310 & \centering 0.8655 & \centering\arraybackslash 0.7624 \\
 \hline 

\end{tabular}

\begin{tabular}{|p{0.15\textwidth}|p{0.1\textwidth}|p{0.1\textwidth}|p{0.1\textwidth}|}
    \hline
    \centering Avg-AAE &  \centering\arraybackslash 0.9994 & \centering 0.9964  & \centering\arraybackslash 0.9910  \\
    \hline
\end{tabular}

\caption{Performance evaluation on larger quantum states (6-qubit and 8-qubit). The last separate row shows the results of AAE for comparison.}
\label{tab:larger_states}
        
\end{table}

\subsection{Impact of Model Architecture}

\label{ssec:impact_model_arch}

% As an initial exploration, the best model architecture of \name{} remains to be further explored.
As a preliminary investigation, the optimal model architecture for \name{} still requires further exploration. 
Currently, we have set the size of the hidden units at a constant 512 (Table~\ref{tab:mlp_config}). 
However, as the number of qubits, $n$, increases, a wider network architecture may become necessary.
% For example, 
% , we set the size of hidden units to be fixed at 512.
% However, we may need wider network architecture with the increasing of $n$.
To showcase the impact of model width, 
we adjust the size to $4\times 2^n$ for 6-qubit states and $16\times 2^n$ for 8-qubit states,
and compare their performance with the original settings,
as shown in Table~\ref{tab:impact_hidden_size}.
% we set the size to be $4\times 2^n$ for 6-qubit states, and $16\times 2^n$ for 8-qubit states the compare their performance with original settings (Table~\ref{tab:impact_hidden_size}).
% As one can observe, this simple change effectively increases the fidelity of \name{},
% indicating the potentially large room to improve the performance of \name{} by designing dedicated network architecture.
As evident from the results, this simple adjustment significantly enhances the fidelity of \name{}, 
suggesting that there is substantial potential to boost \name{}'s performance by developing a more tailored network architecture.

\begin{table}[!h]
        \centering
        
\begin{tabular}{|p{0.2\textwidth}|p{0.12\textwidth}|p{0.12\textwidth}|p{0.12\textwidth}|p{0.12\textwidth}|}
\hline 
\rowcolor{lightgray}
  & \multicolumn{2}{c|}{$n=6$} & \multicolumn{2}{c|}{$n=8$} \\
\hline 
\rowcolor{lightgray}
\centering Dataset
 & 
\centering $\displaystyle h=512$
 & 
\centering $\displaystyle h=4\times 2^{6}$
 & 
\centering $\displaystyle h=512$
 & 
\centering\arraybackslash $\displaystyle h=16\times 2^{8}$
 \\
\hline 
\centering Uniform
 & 
\centering 0.9254
 & 
\centering \textbf{0.9267}
 & 
\centering 0.8648
 & 
\centering\arraybackslash \textbf{0.8821}
 \\
\hline 
\centering Normal
 & 
 \centering 0.7457
 & 
 \centering \textbf{0.7580}
 & 
 \centering 0.6075
 & 
 \centering\arraybackslash\textbf{0.6401}
 \\
\hline 
 
\centering Log-normal
 & 
\centering 0.8575
 & 
\centering \textbf{0.8608}
 & 
\centering 0.7122
 & 
\centering\arraybackslash \textbf{0.7294}
 \\
\hline 
\centering Exponential
 & 
\centering \textbf{0.8757}
 & 
\centering 0.8732
 & 
\centering 0.7613
 & 
\centering\arraybackslash \textbf{0.7781}
 \\
\hline 
\centering Dirichlet
 & 
\centering 0.9232
 & 
\centering \textbf{0.9261}
 & 
\centering 0.8663
 & 
\centering\arraybackslash \textbf{0.8805}
 \\
\hline 
\centering Avg
 & 
\centering 0.8655
 & 
\centering \textbf{0.8690}
 & 
\centering 0.7624
 & 
\centering\arraybackslash \textbf{0.7820}
 \\
 \hline
\end{tabular}

\caption{Impact of increasing network width. Here $h$ refers to the size of hidden units.}
\label{tab:impact_hidden_size}
        
\end{table}

\subsection{Impact of Training Datasets}

\label{ssec:impact_training_dataset}

% Besides model architecture, we may also need a specially designed dataset for pre-training of \name{}.
In addition to refining the model architecture, the development of a specially designed dataset for pre-training \name{} is essential.
% The currently used dataset is FractalDB~\cite{kataoka2020fractal_db},
Currently, the dataset utilized is FractalDB~\cite{kataoka2020fractal_db},
which is originally designed for computer vision tasks.
However, given the wide range of applications of QSP, there is a need to accommodate diverse types of classical data from various domains.
% However, QSP has broad application scenarios, requiring to load various types of classical data from different domains into quantum states.
Therefore, how to create a comprehensive dataset that could fully unleash the potential of \name{} remains an open question.
While developing a pre-trained model that performs well in all kinds of applications may be challenging,
we advocate for a strategy that combines pre-training with fine-tuning for the practical deployment of \name{}, 
similar to the approach used with foundation models in classical machine learning.
% we advocate for a methodology of pre-training plus fine-tuning, for practical deployment of \name{},
% akin to the usage of foundation models in classical domains.
To substantiate this approach, we have compiled a separate dataset that encompasses a variety of statistical distributions not limited to those utilized for evaluation (but with different settings). 
As demonstrated in Table~\ref{tab:impact_finetune}, after fine-tuning, the performance of \name{} improves by approximately 0.03. 
% This enhancement underscores the effectiveness of our proposed methodology in optimizing the utility of \name{} across diverse quantum computing applications.
% To support this perspective, we create a separate dataset consisting of a variety of statistical distributions, including but not limited to those used for evaluation (with different settings).
% As listed in Table~\ref{tab:impact_finetune}, after fine-tuning, the performance of \name{} has improved by $\sim$0.03.

\begin{table}[!h]
        \centering
        
\begin{tabular}{|p{0.2\textwidth}|p{0.15\textwidth}|p{0.3\textwidth}|}
\hline 
\rowcolor{lightgray}
 \centering Dataset & \centering Pre-training & \centering\arraybackslash Pre-training$\displaystyle +$Finetuning \\
\hline 
 \centering Uniform & \centering 0.9731 & \centering\arraybackslash \textbf{0.9909} \\ \hline
 \centering Normal & \centering 0.8201 & \centering\arraybackslash \textbf{0.8879} \\
\hline 
 \centering Log-normal & \centering 0.9421 & \centering\arraybackslash\textbf{0.9717} \\
\hline 
 \centering Exponential & \centering 0.9464 & \centering\arraybackslash\textbf{0.9729} \\
\hline 
 \centering Dirichlet & \centering 0.9737 & \centering\arraybackslash\textbf{0.9903} \\
\hline 
 \centering Avg & \centering 0.9310 & \centering\arraybackslash\textbf{0.9627} \\
 \hline
\end{tabular}

% \begin{tabular}{|p{2cm}|p{3cm}|p{4cm}|}
% \hline
% \centering Column 1 & \centering Column 2 & \centering\arraybackslash Column 3 \\ \hline
% \centering Data 1 & \centering Data 2 & \centering\arraybackslash Data 3 \\ \hline
% \end{tabular}

\caption{Fidelity improvements after fine-tuning \name{} using a dataset consisting of different distributions.}
\label{tab:impact_finetune}
        
\end{table}

% \subsection{Feasibility in Noisy Environments}

% Optionally include supplemental material (complete proofs, additional experiments and plots) in appendix.
% All such materials \textbf{SHOULD be included in the main submission.}

\section{Runtime Estimation for Training on Real Devices}
\label{sec:resource_est_real_device}

Although we have theoretically analyzed the feasibility of training \name{} using states from real devices (Section~\ref{ssec:design}), 
its practical implementation poses significant challenges. 
Specifically, state-of-the-art quantum tomography techniques, such as classical shadow~\cite{zhang2021experimental_classical_shadow,huang2022learning_states_classical_shadow}, require numerous \emph{snapshots}, 
each measuring a distinct observable. 

To train \name{}, each sample in the training dataset necessitates one classical shadow to obtain the prepared state. 
For instance, with the FractalDB-60 dataset, one training epoch requires 60,000 classical shadows. 
Our experiments on the IBM cloud platform reveal an average runtime of 3.02 seconds per circuit job excluding queuing time.
Suppose the number of snapshots is 1000,
then the total runtime to train \name{} for 10 epochs is about 1,812,000,000 seconds\footnote{$10\times 1000 \times 60000 \times 3.02$}, roughly 57 years, 
making the process prohibitively expensive and time-consuming.

However, quantum tomography is under active investigation, and we expect more efficient techniques to emerge for acquiring noisy quantum states from real devices. 
Additionally, with the advancement of quantum computing system, future systems may have tightly integrated quantum-classical heterogeneous architectures (shorter runtime per job) while being capable of executing numerous quantum circuits in parallel (jobs within a classical shadow can execute in parallel).
Hence, we anticipate the training of \name{} to be feasible in the future.

\end{document}